\documentclass[10pt]{iopart}
\usepackage{iopams}
\usepackage{hyperref}
\usepackage{graphicx}
\usepackage{caption}
\hypersetup{
colorlinks=true,
linkcolor=blue,
citecolor=blue,
urlcolor=blue}
\newcommand{\expval}[1]{\langle {#1} \rangle}

\usepackage{color,soul}

\expandafter\let\csname equation*\endcsname\relax
\expandafter\let\csname endequation*\endcsname\relax

\usepackage{dsfont}

\usepackage{mathtools}

\DeclarePairedDelimiterX\braket[2]{\langle}{\rangle}{#1 \delimsize\vert #2}

\begin{document}

\title[Generalized quantum geometric tensor for excited states using the path integral approach]
{Generalized quantum geometric tensor for excited states using the path integral approach}

\author{Sergio B. Ju\'arez$^1$, Diego Gonzalez$^{2}$, Daniel Guti\'errez-Ruiz$^1$ and J. David Vergara$^1$}

\address{$^1$Departamento de F\'isica de Altas Energ\'ias, Instituto de Ciencias Nucleares, Universidad Nacional Aut\'onoma de M\'exico, Apartado Postal 70-543, Ciudad de M\'exico, 04510, M\'exico}

\address{$^2$Escuela Superior de Ingenier\'ia Mec\'anica y El\'ectrica - Instituto Polit\'ecnico Nacional (ESIME - IPN), 07738, CDMX, M\'exico}

\eads{\mailto{sergio1@ciencias.unam.mx}, \mailto{dgonzalezv@ipn.mx}, \mailto{daniel.gutierrez@correo.nucleares.unam.mx} and \mailto{vergara@nucleares.unam.mx}}


\begin{abstract}

The quantum geometric tensor, composed of the quantum metric tensor and Berry curvature, fully encodes the parameter space geometry of a physical system. We first provide a formulation of the quantum geometrical tensor in the path integral formalism that can handle both the ground and excited states, making it useful to characterize excited state quantum phase transitions (ESQPT). In this setting, we also generalize the quantum geometric tensor to incorporate variations of the system parameters and the phase-space coordinates. This gives rise to an alternative approach to the quantum covariance matrix, from which we can get information about the quantum entanglement of Gaussian states through tools such as purity and von Neumann entropy.  Second, we demonstrate the equivalence between the formulation of the quantum geometric tensor in the path integral formalism  and other existing methods.  Furthermore, we explore the geometric properties of the generalized quantum metric tensor in depth by calculating the Ricci tensor and scalar curvature for several quantum systems, providing insight into this geometric information.

\end{abstract}


\section{Introduction}
In the classical sense, all fundamental interactions have a geometric description based on connections  as abelian and non-abelian gauge fields in the case of the standard model or Christoffel symbols in the case of general relativity. However, this geometric notion is lost in quantum mechanics, and one speaks about Hilbert spaces, $C^*$ algebras, discrete spectra, scattering amplitudes, quantum phase transitions, Berry phases, and quantum entanglement, just to mention some of the terms used in the quantum realm.   For several years, there have been various attempts to make a geometric description of quantum mechanics; from a mathematical point of view, the most successful seems to be the non-commutative geometry of Connes \cite{Connes}. In this approach, the geometric description starts from a spectral triplet $(\mathcal{A, \ H, \ D})$ given by a $C^*-$algebra $\mathcal{A}$, a Hilbert space $\mathcal{H}$, and a Dirac operator $\mathcal{D}$. Although interesting in many aspects, this construction does not directly apply to most non-relativistic quantum mechanics problems.

In recent years, geometric tools have been introduced in the realm of information theory, as seen in works such as \cite{Amari}. By employing them, one can gain insights into various aspects of statistical theory. Given that quantum mechanics is a statistical theory, it is possible to pursue a similar approach and obtain a geometric description of quantum mechanics, as has been done in \cite{Provost, Marmo1}. It is important to note that statistical distance is independent of quantum mechanics and can be defined in any probability space. As such, we can develop a deeper understanding of quantum mechanics and its statistical foundations by applying geometric ideas. Some examples of these developments can  be found in \cite{Sarkar1, Sarkar2, Sarkar3}. Furthermore, these studies can be extended using numerical methods to understand more realistic systems better \cite{Diego1, Diego2} and offer direct contact with the experiment \cite{exp1, exp2, exp3}.

In a recent publication  \cite{ADV}, using a Lagrangian approach of the path integral formalism, the authors construct a formulation of the quantum geometric tensor that only accounts for the ground state. However, the quantum geometric tensor is well-defined not only for the ground state but for all eigenstates of the Hamiltonian, which has remained as a drawback of this approach until now. On the other hand, a closely related object is the covariance matrix, which can be interpreted as a Fubiny-Study metric associated with translations in phase space \cite{Abe1993} and has been shown to be related to a generalized quantum geometric tensor \cite{Bogar}. However, this connection comes out naturally in the path integral formalism, as we shall see. This article  has two main contributions. First, we formulate the quantum geometric tensor based on the path integral formalism in a Hamiltonian context for all eigenstates of the system. This generalizes the results of \cite{ADV}, allowing a better understanding of  excited state quantum phase transitions (ESQPTs) and their associated geometric properties. Furthermore, we also demonstrate that this formalism can be generalized to accommodate translations in the phase-space coordinates, which can be used to compute the covariance matrix in similar fashion to the quantum geometric tensor, allowing us to see the connection between concepts naturally. Second, we show that this path integral formulation is on equal footing as those introduced by Provost and Vallee \cite{Provost} and Zanardi et al. \cite{Zanardi}.   

The structure of the paper is the following. In Sec. 2, we introduce the new formulation of the quantum geometric tensor for any quantum state using the Hamiltonian path integral formalism, and prove explicitly that this approach is totally equivalent to the formulations of Provost and Valle \cite{Provost}, and Zanardi et al. \cite{Zanardi}. In this section, we also generalize the quantum geometric tensor to include  translations in the phase-space coordinates. In Sec. 3, we consider the first example of our approach using the generalized oscillator, showing how it works and yields the expected results. In Sec. 4, we compute the quantum geometric tensor for a  generalized  oscillator with a linear term and analyze the behavior of the scalar curvature for this system, showing that the parameter space is almost flat for the $m$-th state with $m>100$. In sec 5, we study the parameter space geometry of general Gaussian state with an arbitrary functional dependency on two adiabatic parameters, and compare it with that of a generalized  oscillator with a linear term. In Sec. 6, we study a system of symmetrically coupled harmonic oscillators, and show that the entanglement information of the ground state is contained in the phase space contribution of the generalized quantum geometric tensor. In Sec. 7, we analyze the geometry of the parameter space of a set of linearly coupled harmonic oscillators. We conclude in Sec. 8 with final remarks.

\section{The Quantum Geometric Tensor for any excited state under a path integral approach}

To obtain the geometric quantum tensor of the $n$th quantum state using path integrals, we begin by assuming that a Hamiltonian $H_i$ describes our system during the time interval $(-\infty, 0)$. We also assume that this Hamiltonian depends on a set of phase-space coordinates $\{q^a,p_a\}$ with $a=1,\dots,N$, as well as a set of parameters $\{\lambda_k\}$ with $k=1,\dots,M$. We denote the complete set of phase-space coordinates and parameters by $\{z^A\}=\{q^a,p_a,\lambda_k\}$, where $A=1,\dots,2N+M$.  Furthermore, we consider that $H_i$ possesses a discrete and nondegenerate spectrum $E_n$. Let us also consider that at the time $t=0$, the system suffers a perturbation that modifies the Hamiltonian in the form
\begin{equation}\label{hamfin}
H_i \xrightarrow{t=0} H_f = H_i + \mathcal{O }_A \delta z^A,
\end{equation}
where $ \mathcal{O}_A$ are deformation functions defined by 
\begin{equation}\label{defope}
 \mathcal{O}_A=\frac{\partial H_i}{\partial z^A} ,
\end{equation}
and that the system evolves in the  remaining time interval $(0,\infty)$  with the new Hamiltonian $H_f$. Notice that here we are considering more general variations than those given in \cite{ADV}, since we are also taking into account variations of the phase space variables $\{q^a,p_a\}$. However, these variations must only be translations, so that we can extract these from the quantum averages. It is important to point out that these deformation functions are always obtained by  differentiating the Hamiltonian function and then they can be translated into quantum operators by using the substitution of the variables $ q\to \hat{q}$ and $p\to \hat{p}$, and applying Weyl ordering. Throughout this paper we use natural units with $\hbar=1$.

With this in mind, we can write the probability  amplitude $\left< q_f,t_f| q_i, t_i \right>$ by inserting two identities in the basis of the energy,  $\hat{H}_i |m\rangle =E_m^i|m\rangle $, as
\begin{align}
 \left< q_f,t_f | q_i, t_i \right> = \sum_{n} \left< q_f, t_f | n\right> \left< n | q_i, t_i \right> 
 = \sum_{n,n'} \left< q_f\right| e^{- i t_f E_{n'}^f} \left| n' \right>\langle n' |n\rangle  \left< n \right| e^{it_i E_{n}^i} \left| q_i \right>, \label{ampli1}
\end{align}
where the subindexes $i$ and $f$ refer to the Hamiltonian that describes the theory. Next, by multiplying \eqref{ampli1} by  $e^{+i t_f E_{m'}^f} e^{-i t_i E_{m}^i}$ and assuming that the states are orthonormal, i.e., $\left<n'|n\right>=\delta_{n'n}$, we get  
\begin{align}
&\nonumber e^{+i t_f E_{m'}^f} e^{-i t_i E_{m}^i}\left< q_f,t_f | q_i, t_i \right>  \\ =& e^{+i t_f E_{m'}^f} e^{-i t_i E_{m}^i} \sum_{n,n'} \left< q_f\right| e^{- i t_f E_{n'}^f} \left| n' \right>\langle n' |n\rangle  \left< n \right| e^{it_i E_{n}^i} \left| q_i \right> \nonumber \\ 
=&\langle q_f | e^{- i t_f( E_{0'}^f -E_{m'}^f)} | 0 \rangle \langle 0 |0\rangle  \left< 0 \right| e^{it_i (E_{0}^i-E_m^i)} \left| q_i \right>+ \dots+\langle q_f | m' \rangle \langle m' |m\rangle  \left< m | q_i \right> \nonumber \\
&+\langle q_f | e^{- i t_f( E_{m+1'}^f -E_{m'}^f)} | m'+1 \rangle \langle m'+1 |m+1\rangle  \left< m+1 \right| e^{it_i (E_{m+1}^i-E_m^i)} \left| q_i \right>+\dots \label{ampli2}.
\end{align}

Usually, when regularizing expressions such as \eqref{ampli2}, we introduce a time regularization that keeps the ground energy term. However, we would prefer to keep an excited state in this case. For this, we must notice that since
\begin{equation}
     e^{+i t_i \hat{H}_i}|q_i\rangle=\sum_n e^{+i t_i E_n^i}|n\rangle\left\langle n \mid q_i\right\rangle, 
\end{equation}
then
\begin{equation}
\begin{aligned}
e^{-i t_i E_m^i+i t_i \hat{H}_i}\left|q_i\right\rangle =&\sum_{n=0}^{m-1} e^{i t_i\left(E_n^i-E_m^i\right)}|n\rangle\left\langle n \mid q_i\right\rangle+|m\rangle\left\langle m \mid q_i\right\rangle \\
& +\sum_{n=m+1}^{\infty} e^{i t_i\left(E_n^i-E_m^i\right)}|n\rangle\left\langle n \mid q_i\right\rangle. 
\end{aligned}
\end{equation}
Now, in order to regularize the above expression to only preserve the $m$-th state, we introduce the $i\epsilon$ prescription in the complex plane to the energy of this state \cite{Alvarez2019}, in the form 
\begin{equation}\label{regener}
E_{m'}^f \to  E_{m'}^f+i\epsilon, \ \ \ \  E_{m}^i \to  E_{m}^i+i\epsilon,
\end{equation} 
which is equivalent to a temporal regularization but only for the ground state. In consequence, by taking the limit $t_i\longrightarrow - \infty$, it turns out that all the terms involving states different from the $m$-th state vanish due to the exponentials such as $e^{i t_i\left(E_n^i-E_m^i\right)} \rightarrow  e^{i t_i\left(E_n^i-E_m^i -i\epsilon\right)}\rightarrow0$, and the term that contains the  $m$-th state energy dominates the summation. Resulting in
\begin{equation} \label{8}
\lim_{t_i \to -\infty}   |m\rangle  \left\langle m \mid q_i\right\rangle= \lim_{t_i \to -\infty}  e^{-i t_i (E_{m}^i + i\epsilon) + i t_i \hat{H}_i }\left|q_i\right\rangle =\lim_{t_i \to -\infty}  e^{-i t_i (E_{m}^i + i\epsilon)  }\left|q_i,t_i\right\rangle,
\end{equation}
and using an analogous construction and taking the limit $t_f\longrightarrow + \infty$ we get
\begin{equation} \label{9}
   \lim_{t_f \to +\infty} \left\langle q_f \mid m^\prime\right\rangle\langle m^\prime |  = \lim_{t_f \to +\infty} \langle q_f | e^{+i t_f (E_{m^\prime}^f + i\epsilon) - i t_f \hat{H}_f }= \lim_{t_f \to +\infty}\langle q_f, t_f | e^{+i t_f (E_{m^\prime}^f + i\epsilon) }.
\end{equation}
Notice that the term $\langle q_f | m' \rangle \langle m' |m\rangle  \left< m | q_i \right>$ in \eqref{ampli2} does not contain any exponential function.  Then, by  following this procedure to keep only the $m$-th state in \eqref{ampli2}, we get
\begin{align}
 \left< q_f,\infty | q_i,-\infty \right> & |_{E_{m'}^f \to  E_{m'}^f+i\epsilon, \  E_{m}^i \to  E_{m}^i+i\epsilon} \nonumber\\  &= \langle q_f,\infty | m'\rangle |_{ E_{m'}^f+i\epsilon}  \left< m' | m \right> \left< m | q_i, -\infty \right> |_{ E_{m}^i+i\epsilon } ,\end{align}
and therefore, we have
\begin{equation} \label{eq:traspsants}
\langle m'|m\rangle=\frac{\left. \langle q_f,\infty | q_i,-\infty \rangle \right|_{E_{m'}^f \to  E_{m'}^f+i\epsilon, \  E_{m}^i \to  E_{m}^i+i\epsilon} }{\langle q_f,\infty | m'\rangle |_{ E_{m'}^f+i\epsilon} \left< m | q_i, -\infty \right> |_{ E_{m}^i+i\epsilon }  }.
\end{equation}

 Let us now analyze each term of this expression. The numerator can be rewritten by inserting an identity at the time $t=0$ in the form $\mathbb{I}=\int dq_0 \left| q^0 \right> \left< q^0\right|$. From this, we obtain
\begin{align} \label{eq:trasidcamc}
&\nonumber\left.\left<  q_f, \infty| q_i,-\infty \right> \right|_{E_{m'}^f \to  E_{m'}^f+i\epsilon, \  E_{m}^i \to  E_{m}^i+i\epsilon}  \\ \nonumber&= \int dq^0 \left.\left< q_f,\infty | q^0 \right>\right|_{E_{m'}^f \to  E_{m'}^f+i\epsilon}\left. \left< q^0 | q_i,-\infty \right>\right|_{ E_{m}^i \to  E_{m}^i+i\epsilon}\nonumber \\
&=\left. \int \mathcal{D}q \mathcal{D}p \exp\left(i\int_{-\infty}^{\infty}d\tau  \left(p\dot q  -H_i\right)  -i \int_0^\infty d\tau  \delta z^A \mathcal{O }_A\right)\right|_{ E_{m}^i \to  E_{m}^i+i\epsilon}.
\end{align}
\normalsize
To get the amplitudes $\langle q_f,\infty | m'\rangle |_{ E_{m'}^f+i\epsilon}$  and  $\left< m | q_i, -\infty \right> |_{ E_{m}^i+i\epsilon } $, we first observe that the generating functional of the Green's functions for the ground state is given by \cite{das2019}
\begin{equation}
Z_f [J] = \int \mathcal{D} q  \mathcal{D} p \exp\left(i \int_{-\infty}^\infty d\tau  \left( p\dot q -H_f + J q\right)\right).
\end{equation}
But taking into account our prescription $ E_{m'}^f \to E_{m^\prime}^f + i \epsilon$ and following the same steps that were used to obtain \eqref{8} and \eqref{9}, it is obtained that now the generating functional $Z[J]$ corresponds to the generating function of the $n$-Green's functions of the $m'$-th excited state, {\it i.e.},
\begin{equation}
 \langle m^{ \prime} | \hat q(t_1) \cdots  \hat q(t_n) | m^{ \prime}  \rangle= \frac{1}{i^n Z[0]}\left.\frac{\delta^n Z[J]}{\delta J(t_1)\cdots\delta J(t_n)}\right|_{J=0,\  E_{m'}^f \to  E_{m'}^f+i\epsilon},
  \end{equation}
with the time ordering prescription $t_1>t_2 >\dots >t_n$. Then, assuming that the theory is time-reversal invariant and bearing in mind the prescription $E_{m'}^f \to  E_{m'}^f+i\epsilon$, we get 
\begin{equation}\label{amp1}
\left.\left< q_f,\infty | m' \right> = \sqrt{Z_f}\right|_{E_{m'}^f \to  E_{m'}^f+i\epsilon}.
\end{equation} 
Taking the same considerations in the limit of $t_i \to -\infty$, we obtain
\begin{equation}\label{amp2}
\left.\left< m | q_i,-\infty \right> = \sqrt{Z_i}\right|_{E_{m}^i \to  E_{m}^i+i\epsilon}.
\end{equation}
Then, using \eqref{eq:trasidcamc}, \eqref{amp1}, and \eqref{amp2}, the overlap (\ref{eq:traspsants}) takes the form
\begin{align} \label{eq:trasinttra}
\left.\left< m' | m \right> =   \dfrac{\int \mathcal{D} q \mathcal{D} p\left( e^{ i\int_{-\infty}^\infty d\tau  \left( p\dot q -H_i\right)  } e^{-i\int_{0}^\infty d\tau   \delta z^A \mathcal{O }_A }\right)}{\sqrt{Z_i Z_f}}\right|_{E_{m}^i \to  E_{m}^i+i\epsilon, \ {E_{m'}^f \to  E_{m'}^f+i\epsilon}}.
\end{align} 

To simplify this expression, we introduce the definition of expectation value for an operator $\hat{A}$ in the $m$-th state  as
\begin{equation}\label{valor esperado ops}
\langle \hat{A} \rangle_m=\left<m\right| \hat{A} \left| m \right> =\left. \frac{1}{Z_i} \int \mathcal{D} q \mathcal{D} p\left[ \exp \left( i \int_{-\infty}^\infty d\tau  \left( p\dot q -H_i\right)  \right) A(q) \right]\right|_{E_{m}^i \to  E_{m}^i+i\epsilon} .
\end{equation}
Notice that we are defining the expectation value with respect to the unperturbed system $H_i$.  In consequence the overlap of the $m$-th states takes the form
\begin{equation}\label{overlap ms iepsilon}
\left< m' | m \right>=\left.\frac{\left< e^{-i\int_0^\infty d\tau  \delta z^A \mathcal{\hat{O}}_A}\right>_m}{\sqrt{\frac{Z_f}{Z_i}}} \right|_{E_{m}^i \to  E_{m}^i+i\epsilon}=\left. \frac{\left< e^{-i\int_0^\infty d\tau  \delta z^A \mathcal{\hat{O}}_A}\right>_m}{\sqrt{\frac{\int \mathcal{D} q  \mathcal{D} p e^{i  \int_{-\infty}^\infty d\tau (p\dot q -H_i - \delta z^A \mathcal{O }_A)}}{Z_i}}}\right|_{E_{m}^i \to  E_{m}^i+i\epsilon},
\end{equation}
where we have absorbed the information regarding the $m^\prime$ state in the expectation value of the operator $e^{-i\int_0^\infty d\tau  \delta z^A \mathcal{\hat{O}}_A}$, meaning that all the information regarding the perturbation that occurred at $t=0$ is now contained in the deformation operators $\hat{\mathcal{O}}_A$. Notice that for the cases where $\mathcal{O}_A$ is not inside a path integral, it is always referred as an operator. This simplifies the notation as we do not need to keep note of the $i\epsilon$ prescription, although it is always implied. With this in mind, we write \eqref{overlap ms iepsilon} as
\begin{equation} \label{eq:trasindtra}
\left< m' | m \right> =\dfrac{\left< \exp\left( -i \int_0^\infty d\tau  \delta z^A \mathcal{\hat{O}}_A(\tau)\right)  \right>_m}{\sqrt{\left< \exp \left(-i \int_{-\infty}^\infty d \tau  \delta z^C \mathcal{\hat{O}}_C(\tau) \right) \right>_m}},
\end{equation}
and in consequence we get
\begin{equation}\label{eq-overlap}
\left| \left< m'\right| \left. m \right>\right|^2= \dfrac{\left\langle e^{\left(-i \int_{0}^\infty d \tau  \delta z^A \mathcal{\hat{O}}_A(\tau) \right)} \right\rangle_m \left\langle e^{\left( -i\int_{- \infty}^0 d \tau  \delta z^B \mathcal{\hat{O}}_B(\tau) \right)}\right\rangle_m}{\left\langle \exp \left(-i \int_{- \infty}^\infty d \tau  \delta z^C \mathcal{\hat{O}}_C(\tau) \right) \right\rangle_m} .
\end{equation}

Expanding \eqref{eq-overlap} in series of the variations $\delta z^A$,  and taking into account that the only allowed variations for the phase space variables $\{q^a, p_a\}$ are translations, we find
\begin{align}\label{eq-overlap2}
&\left| \left< m' | m \right>\right|^2 = 1 +\dfrac{1}{2}\left[  \int_{0}^\infty d\tau_1 \int_{0}^\infty d \tau_2  \left< \mathcal{\hat{O}}_A(\tau_1) \mathcal{\hat{O}}_B (\tau_2) \right>_m + \int_{-\infty}^0 d\tau_1 \int_{-\infty}^0 d \tau_2  \left< \mathcal{\hat{O}}_A(\tau_1) \mathcal{\hat{O}}_B (\tau_2) \right>_m  \right. \nonumber \\& -  \int_{-\infty}^\infty d\tau_1 \int_{-\infty}^\infty d \tau_2  \left< \mathcal{\hat{O}}_A(\tau_1) \mathcal{\hat{O}}_B (\tau_2) \right>_m   \left. + 2\int_{0}^\infty d\tau_1 \int_{-\infty}^0 d \tau_2  \left< \mathcal{\hat{O}}_A(\tau_1) \right>_m \left< \mathcal{\hat{O}}_B (\tau_2) \right>_m \right] \delta z^A \delta z^B .
\end{align}
Here the linear term vanishes. The next step to simplify this expression is to notice that
\begin{align}
\int_{-\infty}^\infty d\tau_1 \int_{-\infty}^\infty d \tau_2 f(\tau_1,\tau_2
) &=& \int_{-\infty}^0 d\tau_1 \int_{-\infty}^0 d \tau_2 f(\tau_1,\tau_2
) +\int_{-\infty}^0 d\tau_1 \int_{0}^\infty d \tau_2 f(\tau_1,\tau_2
) \nonumber \\ & &+\int_{0}^\infty d\tau_1 \int_{-\infty}^0 d \tau_2 f(\tau_1,\tau_2
) + \int_{0}^\infty d\tau_1 \int_{0}^\infty d \tau_2 f(\tau_1,\tau_2
).
\end{align}
and that the time-reversal symmetry of the system state implies
\begin{equation}
\left<  \mathcal{\hat{O}}_A(\tau_1) \mathcal{\hat{O}}_B (\tau_2)  \right>_m =  \left<  \mathcal{\hat{O}}_A(-\tau_1) \mathcal{\hat{O}}_B (-\tau_2)  \right>_m.
\end{equation}
Bearing in mind these facts, \eqref{eq-overlap2} reduces to
\begin{align}\label{overlap2}
\left| \left< m' | m\right>\right|^2 = 1 -  \int_{-\infty}^0 d\tau_1 \int_{0}^\infty d \tau_2  \bigg[ & \left<  \mathcal{\hat{O}}_A(\tau_1) \mathcal{\hat{O}}_B (\tau_2)  \right>_m   \nonumber \\ &
- \left<   \mathcal{\hat{O}}_A(\tau_1) \right>_m \left<  \mathcal{\hat{O}}_B (\tau_2)  \right>_m \bigg]\delta z^A \delta z^B.
\end{align}

Now, the quantum fidelity for the pure states $\left| m\right>$ and $\left| m'\right>$ is defined by \cite{Chrus,Gu}
\begin{equation}\label{fidelity}
F(z,z+\delta z) := \left| \left< m' | m\right>\right|.
\end{equation}
Then, expanding the fidelity to the lowest order, it is natural to introduce the corresponding quantum geometric tensor $G^{(m)}_{AB}$ as
\begin{equation}\label{fidelity2}
F(z,z+\delta z) = 1  +  \frac{1}{2}G^{^{(m)}}_{AB}\delta z^A  \delta z^B,
\end{equation}
By using \eqref{overlap2} and \eqref{fidelity}, we can read off from \eqref{fidelity2} the quantum geometric tensor $G^{(m)}_{AB}$ for the $m$-th state,
\begin{equation}\label{eq:defQGT}
G_{AB}^{(m)}  =   -\int_{-\infty}^0 d\tau_1 \int_0^{\infty} d \tau_2 \left[  \left<  \mathcal{\hat{O}}_A(\tau_1) \mathcal{\hat{O}}_B(\tau_2) \right>_m \right.  \left. - \left< \mathcal{\hat{O}}_A(\tau_1)  \right>_m \left< \mathcal{\hat{O}}_B(\tau_2) \right>_m \right].
\end{equation}
Notice that from this generalized quantum geometric tensor we can obtain both the regular quantum geometric tensor  associated to variations of the parameters $\{\lambda\}$ and a quantum geometric tensor for associated to translational variations of the phase-space coordinates. It is important to point out that the generalized quantum geometric tensor \eqref{eq:defQGT} was first presented without derivation in \cite{Bogar}; here we have derived it by using path integrals. From the real part of \eqref{eq:defQGT} we get  the Fubini-Study metric of the $m$th quantum state 
\begin{align} \label{eq:tmqintdetra}
\textbf{Re} G_{AB}^{(m)}  =   \textbf{Re} \left(-\int_{-\infty}^0 d\tau_1 \int_0^{\infty} d \tau_2 \bigg[ \right.   \left< \mathcal{\hat{O}}_A(\tau_1) \mathcal{\hat{O}}_B(\tau_2)\right> \left._m -\left. \left< \mathcal{\hat{O}}_A(\tau_1) \right> \right._m \left. \left< \mathcal{\hat{O}}_B(\tau_2)\right> \right._m \bigg]\right)  \nonumber \\ 
= -\int_{-\infty}^0 d\tau_1 \int_0^{\infty} d \tau_2 \left[ \frac{1}{2}\left. \left< \left\lbrace \mathcal{\hat{O}}_A(\tau_1), \mathcal{\hat{O}}_B(\tau_2) \right\rbrace_+ \right>\right._m  -\left. \left< \mathcal{\hat{O}}_A(\tau_1) \right>\right._m \left. \left< \mathcal{\hat{O}}_B(\tau_2)\right>\right._m \right] ,
\end{align} 
which is a generalization of the quantum metric tensor, whereas from the imaginary part we obtain a generalization of the Berry curvature, namely 
\begin{align}
\textbf{Im} G_{AB}^{(m)} &= \frac{-1}{2 i}  \int_{-\infty}^0 d\tau_1 \int_0^{\infty} d \tau_2  \left( \left. \left< \mathcal{\hat{O}}_A (\tau_1) \mathcal{\hat{O}}_B(\tau_2)\right>\right._m -   \left.\left< \mathcal{\hat{O}}_A(\tau_1) \mathcal{\hat{O}}_B(\tau_2)\right>\right._m  \right) \nonumber \\  &= \frac{-1}{2 i}  \int_{-\infty}^0 d\tau_1 \int_0^{\infty} d \tau_2 \left. \left< \left[ \mathcal{\hat{O}}_A(\tau_1), \mathcal{\hat{O}}_B(\tau_2) \right]\right>\right._m. 
\end{align}

In summary, we have  generalizations for the quantum metric tensor for the $m$th quantum state, namely
\begin{equation} \label{eq:tmcdc}
g_{AB}^{(m)} = -\int_{-\infty}^0 d\tau_1 \int_0^{\infty} d \tau_2 \left[ \frac{1}{2} \left. \left< \left\lbrace \mathcal{\hat{O}}_A(\tau_1), \mathcal{\hat{O}}_B(\tau_2) \right\rbrace_+\right>\right._m - \left.\left< \mathcal{\hat{O}}_A(\tau_1) \right>\right._m \left. \left< \mathcal{\hat{O}}_B(\tau_2)\right> \right._m \right],
\end{equation}
and for the Berry curvature, which in general is $F_{AB}^{(m)}=-2 \textbf{Im} G_{AB}^{(m)}$,
\begin{equation} \label{eq:cbdc}
F_{AB}^{(m)} =  \frac{1}{ i}  \int_{-\infty}^0 d\tau_1 \int_0^{\infty} d \tau_2  \left.\left< \left[ \mathcal{\hat{O}}_A(\tau_1), \mathcal{\hat{O}}_B(\tau_2) \right]\right>\right._m,
\end{equation}
defined for arbitrary variations of the parameters and translational variations of the phase space variables. It is worth mentioning that even if all the above expressions are written in terms of operators, it is also always possible to translate them in terms of path integrals using \eqref{valor esperado ops}.\newline

It is important to remark that the metric \eqref{eq:tmcdc} and the curvature \eqref{eq:cbdc} extend the ones previously reported in \cite{ADV} in two ways. First, these expressions are valid for any excited quantum state, not just for the ground state. Second, they take into account variations of the translation type for the phase space variables, as well as of the system parameters. Therefore, they further extend the results reported in \cite{Provost}, and also incorporate the metric introduced by \cite{Abe1993}. As a result, the metric tensor \eqref{eq:tmcdc} and curvature \eqref{eq:cbdc} contain more information than the usual quantum metric tensor and Berry curvature, respectively.

\subsection{Quantum covariance matrix, purity and von Neumann entropy}

The  quantum  metric tensor for the phase space is intimately related to the quantum covariance matrix $\sigma = (\sigma_{\alpha \beta})$ of a quantum state $|m\rangle$. More precisely, the entries $(\sigma_{\alpha \beta})$ of the quantum covariance matrix are given by
\begin{align}
\sigma_{\alpha \beta} := \frac{1}{2} \langle \hat{r}_{\alpha} \hat{r}_{\beta} + \hat{r}_{\beta} \hat{r}_{\alpha}  \rangle_{m}-\langle \hat{r}_{\alpha}\rangle_{m} \langle \hat{r}_{\beta} \rangle_{m}\,, \label{qumet}
\end{align}
with $\hat{\mathbf{r}}=(\hat{q}_1,\dots,\hat{q}_N,\hat{p}_1,\dots,\hat{p}_N)^T$ and $\alpha,\beta = 1,\dots, 2N$. This matrix can be related to the components of the phase-space part of quantum metric tensor in the following way \cite{Bogar}:
\begin{equation}\label{qcov qgt 1}
 g_{q_{a}q_{b}}^{(m)}= \frac{1}{2}\langle \hat{ {p}}_{a} \hat{ {p}}_{b}+\hat{ {p}}_{b} \hat{ {p}}_{a} \rangle_{m} - \langle \hat{ {p}}_{a} \rangle_{m} \langle \hat{ {p}}_{b} \rangle_{m}  =  \sigma_{p_{a}p_{b}} ,
\end{equation}
\begin{equation}\label{qcov qgt 2}
 g_{q_{a}p_{b}}^{(m)}=-  \frac{1}{2}\langle \hat{ {p}}_{a} \hat{ {q}}_{b}+\hat{ {q}}_{b} \hat{ {p}}_{a} \rangle_{m} + \langle \hat{ {p}}_{a} \rangle_{m} \langle \hat{ {q}}_{b} \rangle_{m}   = -\sigma_{q_{a}p_{b}} ,
\end{equation}
\begin{equation}\label{qcov qgt 3}
g_{p_{a}p_{b}}^{(m)}= \frac{1}{2}\langle \hat{ {q}}_{a} \hat{ {q}}_{b}+\hat{ {q}}_{b} \hat{ {q}}_{a} \rangle_{m} - \langle \hat{ {q}}_{a} \rangle_{m} \langle \hat{ {q}}_{b} \rangle_{m}, = \sigma_{q_{a}q_{b}},
\end{equation}
where $a,b=1,\dots,N$ stand for the degrees of freedom. Moreover, the imaginary part of the phase space quantum geometric tensor is related to the symplectic structure as
\begin{equation}
    F_{\alpha \beta} ^{(m)} = -\Omega_{\alpha \beta} = i [ \hat{r}_\alpha, \hat{r}_\beta  ],
\end{equation}
where $\Omega_{\alpha \beta}$ is the symplectic matrix \cite{Bogar}.
Therefore, by deriving the generalized quantum geometric tensor, we have laid the groundwork for this alternative approach to compute both the quantum covariance matrix and the symplectic structure. 

 On the other hand, for Gaussian states, all the information needed to construct the purity and the von Neumann entropy is contained within the quantum covariance matrix and, thus, within the generalized quantum geometric tensor. Indeed, the purity for a Gaussian state takes the particular form \cite{deGosson, PurityGaussian, Russianpurity}
\begin{equation}\label{purity gaussian}
\mu\left(a_{1}, a_{2}, \ldots, a_{n}\right)=\left(\frac{1}{2}\right)^{n} \frac{1}{\sqrt{\operatorname{det} \sigma_{(n)}}},
\end{equation}
where $\sigma_{(n)}$ is the reduced covariance matrix of subsystem composed of $a_{1}, a_{2}, \ldots, a_{n}$ particles of the system of $N$ degrees of freedom.

For the von Neumann entropy, which is another measure of entanglement, in the case of Gaussian states, can easily be computed by getting the symplectic eigenvalues of the corresponding reduced covariance matrix. In this sense, we can write the von Neumann entropy for a Gaussian state as \cite{deGosson,serafini2017quantum,Demarie_2018}
\begin{equation}\label{von Neumann OASA}
S_V\left(a_{1}, a_{2}, \ldots, a_{n}\right)=\sum_{k=1}^{n} \mathcal{S}\left(\nu_{k}\right),
\end{equation}
with
\begin{equation}\label{vN symplectic}
\mathcal{S}\left(\nu_{k}\right)=\left(\nu_{k}+\frac{1}{2}\right) \ln \left(\nu_{k}+\frac{1}{2}\right)-\left(\nu_{k}-\frac{1}{2}\right) \ln \left(\nu_{k}-\frac{1}{2}\right),
\end{equation}
where $\nu_{k}$ are the symplectic eigenvalues of the reduced quantum covariance matrix $\sigma_{(n)}$ of the subsystem. Notice that $S\left(\nu_{k}\right)=0$ only if $\nu_{k}=\frac{1}{2}$ \cite{deGosson}.\newline

 In the next subsection, we will prove our formalism's equivalence to the literature results previously reported in \cite{Abe1993,Zanardi,Chrus}.
 
\subsection{Alternative form of the generalized quantum geometric tensor}

The parameter part of the generalized quantum geometric tensor \eqref{eq:defQGT}  seems to be different from the ones previously reported \cite{Provost,Zanardi,Gu}, in particular since it only requires information from the $m$-th state. For example, the  formulation of the quantum metric tensor first reported by Zanardi et al. in \cite{Zanardi} requires information from all quantum states for its computation. Here, we show that all the expressions are completely equivalent following the approach of \cite{Bogar}. 

Starting  from \eqref{eq:defQGT}, which we take to the Schrödinger picture where the operators are time-independent. This means that if we have a generic Heisenberg
operator $\hat{{\cal O}}(t)$, we can write $\hat{{\cal O}}(t)=e^{i\hat{H}t}\hat{{\cal O}}e^{-i\hat{H}t}$ with $\hat{{\cal O}}$ a Schrödinger operator $\hat{{\cal O}}(t=0)$. Bearing this in mind, the second term of \eqref{eq:defQGT} becomes $\langle\hat{{\cal O}}_{A}(t_{1})\rangle_{n}=\langle n|e^{i\hat{H}t_{1}}\hat{{\cal O}}_{A}e^{-i\hat{H}t_{1}}|n\rangle=\langle n|\hat{{\cal O}}_{A}|n\rangle$, since we can apply the first exponential to the left and the second to the right, canceling each other out. Now, using the same manipulation,  the first term of \eqref{eq:defQGT} takes the form
\begin{equation}
\langle\hat{{\cal O}}_{A}(t_{1})\hat{{\cal O}}_{B}(t_{2})\rangle_{n}=e^{-iE_{n}(t_{2}-t_{1})}\langle n|\hat{{\cal O}}_{A}e^{-i\hat{H}t_{1}}e^{i\hat{H}t_{2}}\hat{{\cal O}}_{B}|n\rangle.
\end{equation}
Then, inserting the identity operator $\mathbb{I}=\sum_{m}|m\rangle\langle m|$
between the two exponentials inside the bracket, we find
\begin{equation}
\langle\hat{{\cal O}}_{A}(t_{1})\hat{{\cal O}}_{B}(t_{2})\rangle_{n}=\sum_{m}e^{i(E_{m}-E_{n})(t_{2}-t_{1})}\langle n|\hat{{\cal O}}_{A}|m\rangle\langle m|\hat{{\cal O}}_{B}|n\rangle.
\end{equation}
The sum in this expression runs over all the allowed values of the quantum number $m$ and can be split into two terms as 
\begin{align}
\langle\hat{{\cal O}}_{A}(t_{1})\hat{{\cal O}}_{B}(t_{2})\rangle_{n}=  \sum_{m\neq n}e^{i(E_{m}-E_{n})(t_{2}-t_{1})}\langle n|\hat{{\cal O}}_{A}|m\rangle\langle m|\hat{{\cal O}}_{B}|n\rangle+\langle n|\hat{{\cal O}}_{A}|n\rangle\langle n|\hat{{\cal O}}_{B}|n\rangle.
\end{align}

From these results, it is now easy to notice that the integrand in \eqref{eq:defQGT} reduces to
\begin{equation}
\langle\hat{{\cal O}}_{A}(t_{1})\hat{{\cal O}}_{B}(t_{2})\rangle_{n}-\langle\hat{{\cal O}}_{A}(t_{1})\rangle_{n}\langle\hat{{\cal O}}_{B}(t_{2})\rangle_{n}=\sum_{m\neq n}e^{i(E_{m}-E_{n})(t_{2}-t_{1})}\langle n|\hat{{\cal O}}_{A}|m\rangle\langle m|\hat{{\cal O}}_{B}|n\rangle,
\end{equation}
and hence,
\begin{equation}\label{QTMtime}
G_{AB}^{(n)}=- \sum_{m\neq n}\left(\intop_{-\infty}^{0}\mathrm{d}t_{1}\intop_{0}^{\infty}\mathrm{d}t_{2}\,e^{i(E_{m}-E_{n})(t_{2}-t_{1})}\right)\langle n|\hat{{\cal O}}_{A}|m\rangle\langle m|\hat{{\cal O}}_{B}|n\rangle.
\end{equation}
Notice that we have completely isolated the time-dependence and that we can get rid of it by performing the integrals. In fact, having in mind
the ranges of $t_{1}$ and $t_{2}$, we have
\begin{equation}
\intop_{-\infty}^{0}\mathrm{d}t_{1}\intop_{0}^{\infty}\mathrm{d}t_{2}\,e^{i(E_{m}-E_{n})(t_{2}-t_{1})}=\lim_{\epsilon\rightarrow0^{+}}\intop_{-\infty}^{0}\mathrm{d}t_{1}\intop_{0}^{\infty}\mathrm{d}t_{2}\,e^{i(E_{m}-E_{n}+i\epsilon)(t_{2}-t_{1})}=-\frac{1}{(E_{m}-E_{n})^{2}},
\end{equation}
which together with \eqref{QTMtime} yields \cite{Bogar}
\begin{equation}\label{QGTpert}
G_{AB}^{(n)}=\sum_{m\neq n}\frac{\langle n|\hat{O}_{A}|m\rangle\langle m|\hat{O}_{B}|n\rangle}{(E_{m}-E_{n})^{2}}.
\end{equation}
This is the Schrödinger form of the generalized quantum geometric tensor \eqref{eq:defQGT}, from which it is also straightforward to compute the generalized quantum metric tensor and Berry curvature.

We can see that if we only consider the variations on the parameter space in \eqref{QGTpert}, it naturally becomes
\begin{equation}
G_{ij}^{(n)}=\sum_{m\neq n}\frac{\langle n|\partial_{i}\hat{H}|m\rangle\langle m|\partial_{j}\hat{H}|n\rangle}{(E_{m}-E_{n})^{2}},\label{eq:perturbativeform}
\end{equation}
which is the known result from \cite{Zanardi,Carollo2020}, sometimes called the perturbative form of the quantum geometric tensor. One advantage of this formulation is that divergences become explicitly visible when energy level crossing occurs, which is a hallmark of a quantum phase transition \cite{Sachdev}.  However, a major drawback of this formulation is that we must now calculate the Hamiltonian deformations' expectation values over all the system's quantum states, which is generally not an easy task.

From (\ref{eq:perturbativeform}), it is straightforward to arrive
at the form of the quantum geometric tensor originally obtained by Provost and Vallee \cite{Provost}. Differentiating Schrödinger's equation, $\hat{H}|m\rangle=E_{m}|m\rangle$, we note that $\langle\partial_{i}m|n\rangle=\frac{\langle n|\partial_{i}\hat{H}|m\rangle}{E_{n}-E_{m}}$ for $m\neq n$, which substituted in \eqref{eq:perturbativeform} leads to
\begin{equation}
G_{ij}^{(n)}=\sum_{m\neq n}\langle\partial_{i}n|m\rangle\langle m|\partial_{j}n\rangle=\sum_{m}\langle\partial_{i}n|m\rangle\langle m|\partial_{j}n\rangle-\langle\partial_{i}n|n\rangle\langle n|\partial_{j}n\rangle,
\end{equation}
where in the last equality we have added and subtracted one term to complete the sum. In
this expression, we readily recognize the identity operator $\mathbb{I}=\sum_{m}|m\rangle\langle m|$, and hence,
\begin{equation}
G_{ij}^{(n)}=\langle\partial_{i}n|\partial_{j}n\rangle-\langle\partial_{i}n|n\rangle\langle n|\partial_{j}n\rangle.
\end{equation}

This is the most common form of the  quantum geometric tensor \cite{Chrus}, whose real part is the  quantum  metric tensor (or quantum information  metric), $g_{ij}^{(n)}=\mathrm{Re}G_{ij}^{(n)}$, and its imaginary
part is proportional to the Berry curvature, $F_{ij}^{(n)}=-2\mathrm{Im}G_{ij}^{(n)}$. In the next sections, we will consider several representative examples of the computation of the generalized quantum geometric tensor using our formalism (\ref{eq:defQGT}).

\section{Generalized oscillator}

This first system is fundamental since it has a non vanishing  quantum  metric tensor and Berry Curvature. Furthermore, by manipulating its parameters we can study quantum phase transitions or pose it as a damped harmonic oscillator while still being quickly reducible to a regular harmonic oscillator. The Hamiltonian for this system is given by
\begin{equation}\label{GHO}
	H=\frac{1}{2}[Zp^2+Y(pq+ qp)+Xq^2],
\end{equation}
where $\{\lambda_i\} = \{X, Y,Z\}$, with $i=1,2,3$, are real numbers acting as adiabatic parameters. Now we can do small variations on the parameter and translations in the phase space as $\{X, Y, Z\} \to \{X+\delta X, Y+\delta Y,Z+\delta Z,\}$ and $\{r_\alpha\}=\{q, p\} \to \{q+ \delta q, p + \delta p\}$, with $\alpha=1,2$.

The deformation functions needed to compute the generalized quantum geometric tensor \eqref{eq:defQGT} are
\begin{subequations}
    \begin{align}
       \mathcal{O}_X (t)&=\frac{\partial H}{\partial X}= \frac{q^2}{2}, \label{opetarorX} 
       \\
       \mathcal{O}_Y (t)&=\frac{\partial H}{\partial Y}=\frac{pq+ qp}{2},\label{opetarorY}
       \\
       \mathcal{O}_Z (t)&= \frac{\partial H}{\partial Z}=\frac{p^2}{2},
    \end{align}
\end{subequations}
\begin{subequations}
    \begin{align}
        \mathcal{O}_q (t)&= \frac{\partial H}{\partial q}=Yp+ Xq,\\
        \mathcal{O}_p (t)&=\frac{\partial H}{\partial p}=Zp + Yq.
    \end{align}
\end{subequations}

To perform the analysis it is convenient to introduce the linear canonical transformation $\hat{q}=Z^{1/2}\hat{Q}$ and $\hat{p}=Z^{-1/2}(\hat{P}-Y\hat{Q})$, and the creation and annihilation operators defined respectively by
\begin{subequations}\label{creannhi ops OAG}
\begin{align}
    \hat{a}^\dagger =\left(\frac{\omega}{2}\right)^{1/2}\hat{Q}+i\left(\frac{1}{2\omega}\right)^{1/2}\hat{P},\\
    \hat{a}=\left(\frac{\omega}{2}\right)^{1/2}\hat{Q}-i\left(\frac{1}{2\omega}\right)^{1/2}\hat{P},
    \end{align}
\end{subequations}
where $\omega:=(XZ-Y^2)^{1/2}$ is the normal frequency of the system.\\

Taking into account the non-zero terms of the generalized quantum geometric tensor for the $n$th quantum state, we can be split it into two subtensors in the following way
\begin{equation}
	G^{(n)}_{AB}=\
	\begin{pmatrix}
		G_{{ij}}^{(n)} & 0  \\
		0 & G_{{\alpha\beta}}^{(n)} \\
	\end{pmatrix},
\end{equation}
where $G_{{ij}}^{(n)}$ is the usual quantum geometric tensor for the parameter sector and $G_{\alpha\beta}^{(n)}$ is the part that  corresponds to the phase-space translations.\\

For example, to obtain the component $G_{XY}^{(n)}$, associated to variations of the parameters $X$ and $Y$,  we need to first compute the expectation values  $\expval{\mathcal{\hat{O}}_X(t_2)\mathcal{\hat{O}}_Y(t_1)}_n$, $\expval{\mathcal{\hat{O}}_X(t)}_n$ and $\expval{\mathcal{\hat{O}}_Y(t)}_n$ involved in \eqref{eq:defQGT}. Using \eqref{opetarorX} and \eqref{opetarorY}, the first one reads
\begin{equation}
	\expval{\mathcal{\hat{O}}_X(t_1)\mathcal{\hat{O}}_Y(t_2)}_n=\frac{1}{4}\langle\hat q^2(t_1)\left( \hat p(t_2)\hat q(t_2)+\hat q(t_2)\hat p(t_2)\right) \rangle_n,
\end{equation}
Note that the operators on the right side of this equation are at different times, and hence they must be tied to the ones in the Schrödinger picture; then, the Heisenberg equations must be used. After these considerations and using the properties of the creation and annihilation operators, we obtain 
\begin{align}\label{2 1 OXOY}
	\expval{\mathcal{\hat{O}}_X(t_1)\mathcal{\hat{O}}_Y(t_2)}_n \nonumber = & \frac{1}{4}\Big[ \Big(\frac{-YZ}{2\omega^2} -\frac{iZ}{2\omega}\Big)e^{2i\omega(t_1-t_2)}(n(n-1))
	 -\Big(\frac{YZ}{2\omega^2}\Big)(2n+1)^2
	\\& + \Big(\frac{-YZ}{2\omega^2} +\frac{iZ}{2\omega}\Big) e^{-2i\omega(t_1-t_2)}(n+1)(n+2)\Big].
\end{align}
By following an analogous procedure, for the other expectation values we get 
\begin{equation}\label{OXOYterm2}
	\expval{\mathcal{\hat{O}}_X(t)}_n=\frac{Z}{4\omega}(2n+1),
\end{equation}
\begin{equation}\label{OXOYterm3}
	\expval{\mathcal{\hat{O}}_Y(t)}_n=\frac{-Y}{2\omega}(2n+1).
\end{equation}
Inserting \eqref{2 1 OXOY}, \eqref{OXOYterm2}, and \eqref{OXOYterm2} into \eqref{eq:defQGT}, and solving the integrals, the component $G_{XY}^{(n)}$ of the quantum metric tensor is
\begin{equation}\label{2 1 GXY}
	G_{XY}^{(n)}=-\frac{2 YZ}{32\omega^4}(n^2+n+1)+\frac{i Z}{8\omega^3}(n+\frac{1}{2}),
\end{equation}
where $n$ are non-negative integers.

The other eight components are obtained along the same lines, which yield the parameter part of the quantum geometric tensor 
\begin{align}\label{5 TGC OAG par}
	G_{ij}^{(n)}=&\begin{pmatrix}
		G_{XX}^{(n)} & G_{XY}^{(n)}  & G_{XZ}^{(n)} \\
		G_{YX}^{(n)} & G_{YY}^{(n)} & G_{YZ}^{(n)} \\
		G_{ZX}^{(n)} & G_{ZY}^{(n)} & G_{ZZ}^{(n)}
	\end{pmatrix} \nonumber \\ 
 =&\frac{(n^2+n+1)}{32\omega^4} \
	\begin{pmatrix}
		Z^2& -2YZ  & -XZ+2Y^2 \\
		-2YZ & 4ZX & -2YX \\
		-XZ+2Y^2 & -2YX & X^2
	\end{pmatrix}
	\nonumber \\ & +i\frac{(n+\frac{1}{2})}{8\omega^3}
	\begin{pmatrix}
		0 & Z  & -Y \\
		-Z & 0 & X \\
		Y & -X & 0
	\end{pmatrix}.
\end{align}
This expression is the well-known result for the $n$th eigenstate of the generalized oscillator, which includes including both the Metric Tensor and the Berry Curvature, see \cite{Chrus}.\\

In the case of the translational evolution in phase-space, we can use the same prescription but with the operators $\mathcal{\hat{O}}_q$ and $\mathcal{\hat{O}}_p$ coming from the deformation functions $\mathcal{O}_p$ and $\mathcal{O}_q$, respectively. The resulting phase-space part of the generalized quantum geometric tensor is 
\begin{equation}\label{5 TGC OAG pq}
	G_{\alpha\beta}^{(n)} = \left(
	\begin{array}{cc}
		G_{qq}^{(n)} & G_{qp}^{(n)} \\
			G_{pq}^{(n)} & G_{pp}^{(n)} \\
	\end{array}
	\right) = \frac{\left(n+\frac{1}{2}\right) }{ \omega}\left(
	\begin{array}{cc}
		X & Y \\
			Y & Z \\
	\end{array}
	\right)
	+\frac{i}{2   }
	\left(
	\begin{array}{cc}
		0 & 1 \\
		-1 & 0 \\
	\end{array}
	\right).
\end{equation}
 Notice that we got this last expression by taking the expectation values of the configuration-space coordinates and the integration removes all dependence on them. In this sense we cannot consider this metric as pertaining the configuration space. However, to study its geometric properties we will consider that it is another parameter-space metric.

After obtaining the tensors \eqref{5 TGC OAG par} and \eqref{5 TGC OAG pq}, whose real parts correspond to metric tensors, it is natural to study the Riemannian geometry of these new spaces, for which we must obtain the scalar curvature associated with such metrics.

Let us begin with the parameter part. Given that the  metric tensor of the parameter space is singular, i.e. $\det(\textbf{Re}\, G_{ij}^{(n)})=0$ \cite{Gonzalez}, we need to focus on a subset of parameters and consider the associated two-dimensional metric. We start by fixing
the parameter $Z=Z_{0}$, where $Z_{0}$ is a constant different from zero, and then eliminating the third row and column of the real part of \eqref{5 TGC OAG par} since the variation is null. The parameters are now $\{\lambda_{i'}\}=\{X,Y\}$ with $i',j',\dots=1,2$, and the corresponding metric of the two-dimensional submanifold is
\begin{equation}\label{metrixZ0}
	g_{i' j'}^{(n)}=\frac{n^{2}+n+1}{32\omega^{4}}\begin{pmatrix}Z_{0}^{2} & -2YZ_{0}\\
		-2YZ_{0} & 4XZ_{0}
	\end{pmatrix},
\end{equation}
which has determinant  given by $det[g_{i' j'}^{(n)}]=-\frac{\left(n^{2}+n+1\right)^{2}Z_{0}^{2}}{256\omega^{6}}$. To calculate the scalar curvature we can use the expression \cite{Diego1},
\begin{equation}R=\frac{1}{\sqrt{g}}\left\{\frac{\partial}{\partial x^{1}}\left[\frac{1}{\sqrt{g}}\left(\frac{g_{12}}{g_{11}} \frac{\partial g_{11}}{\partial x^{2}}-\frac{\partial g_{22}}{\partial x^{1}}\right)\right]+\frac{\partial}{\partial x^{2}}\left[\frac{1}{\sqrt{g}}\left(2 \frac{\partial g_{12}}{\partial x^{1}}-\frac{\partial g_{11}}{\partial x^{2}}-\frac{g_{12}}{g_{11}} \frac{\partial g_{11}}{\partial x^{1}}\right)\right]\right\},\end{equation}
which in the case of \eqref{metrixZ0} leads to 
\begin{equation}\label{Rz0}
R_\lambda=-\frac{16}{n^{2}+n+1}.
\end{equation}
Of course the identity $R_{i' j' k' l'}=\frac{R}{2}\left(g_{i' k' }g_{j' l' }-g_{i' l' }g_{j' k'}\right)$ is fulfilled, as it must be for a two-dimensional manifold. This means that the Riemann tensor has only one independent component and the scalar curvature contains all the information. We can see that the divergence in the determinant of the metric does not appear in the scalar curvature, and therefore the quantum phase transition is parameter dependent. It is interesting to notice that the scalar curvature \eqref{Rz0} is strictly negative and parameter-independent, which indicates that the parameter space is hyperbolic and maximally-symmetric \cite{Carroll}. Furthermore, when large quantum numbers are considered, it approaches zero, then indicating that the effect of having excited states is the reduction of (negative) curvature, which results in a flattening of the surface.



A similar argument for the other two possibilities for fixing one parameter, namely $X=X_{0}$(=constant) or $Y=Y_{0}$(=constant), lead to different metrics restricted to a submanifold of the original parameter space. However, the scalar curvature is the same for every case. Remarkably, the square root of the determinant of each resultant metric is proportional to a component of Berry's curvature. More precisely, considering the ground state for simplicity, the following relation holds:
\begin{equation}
	F_{ij}^{(0)}=-\frac{1}{2}\epsilon_{ijk}\sqrt{\det g_{{i' j'} }^{(0)}[k]},\label{eq:relPalumbo}
\end{equation}
where $\epsilon_{ijk}$ is the Levi-Citiva symbol ($\epsilon_{123}=+1$) and $g_{i' j' }^{(0)}[k]$ is the metric obtained by eliminating
the $k$-th row and column from the real part  of \eqref{5 TGC OAG par}.

A relation analogous to (\ref{eq:relPalumbo}) appears in  \cite{Palumbo} for
the well-known spin 1/2 particle subject to an adiabatically rotating
magnetic field, which possesses a Berry curvature resembling a magnetic monopole (according to the same reference, the relation seems to hold for systems with tensor monopoles). Although in our case we are dealing with a bosonic generalized harmonic oscillator, which is very different in nature from that of \cite{Palumbo}, it can be shown that the curvatures and metrics of both systems are related by the complex parameter transformation
\begin{equation}
	X=B_{1}+iB_{2},\,\,\,\,Y=-iB_{3},\,\,\,\,Z=B_{1}-iB_{2},
\end{equation}
where $(B_{1},B_{2},B_{3})$ are the components of said magnetic field.

Now we will analyze the geometric properties of the parameter dependent phase-space metric tensor \eqref{5 TGC OAG pq}, in the same way that we did with \eqref{metrixZ0}. Taking as coordinates the parameters $X$ and $Y$ we obtain the associated scalar curvature 
\begin{equation}\label{RpqXY}
R_{XY}=-\frac{ \left(4 X^2 Z+8 X Y^2+6 X Z^2+6 Y^2 Z+3 Z^3\right)}{2 (2 n+1) \omega^5}.
\end{equation}
The result remains unchanged regardless of whether we choose the order of the parameters as $\{\lambda_i\}=\{X,Y\}$ or $\{\lambda_i\}=\{Y,X\}$.
\begin{figure}[h!]
  \centering
  \captionsetup{justification=raggedright,
singlelinecheck=false
}
  \includegraphics[width=0.8\textwidth]{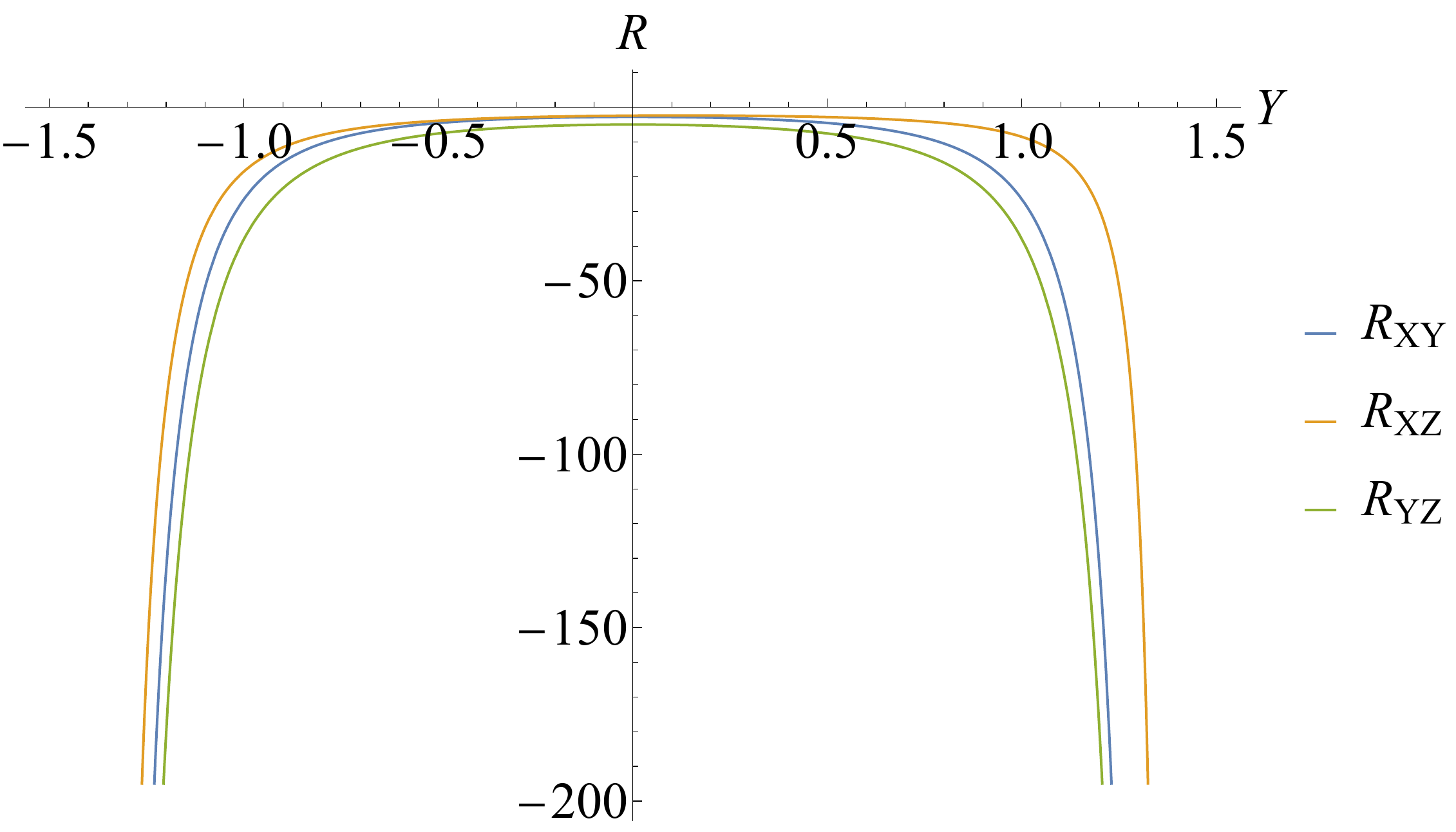}
  \captionof{figure}{Scalar curvatures \eqref{RpqXY}, \eqref{RpqXZ}, and \eqref{RpqYZ} for the parameter dependent phase-space metric tensor as functions of $Y$ with $n=0$, $Z=1,$ and $X=2$.} 
  \label{fig:OAGescalarespq}
\end{figure}

Notice that there is a big difference between the  scalar curvatures \eqref{Rz0} and \eqref{RpqXY}. While \eqref{Rz0} is constant and independent of any parameter, which makes sense since the ground state is Gaussian and for any Gaussian state the scalar curvature is just a constant number (the proof of this statement can be found at the end of the next example), the scalar curvature \eqref{RpqXY} does indeed depend on the parameters. On the other hand they share that when $n\to \infty$, $R\to 0$.

Furthermore if instead we take as variables $X$ and $Z$ we get
\begin{equation}\label{RpqXZ}
R_{XZ}=\frac{ -3 X^3+4 X Y Z+2 Y^3-3 Z^3}{2 (2 n+1) \omega^5},
\end{equation}
and with $Y$ and $Z$
\begin{equation}\label{RpqYZ}
R_{YZ}=-\frac{ 3 X^3+6 X^2 Z+6 X Y^2+4 X Z^2+8 Y^2 Z}{2 (2 n+1) \omega^5}.
\end{equation}  
Therefore, the scalar curvature of this space depends on the parameters that we take as variables, in contrast with \eqref{Rz0} which stayed the same for every option.  

A key property of the scalar curvatures \eqref{RpqXY}, \eqref{RpqXZ}, and \eqref{RpqYZ} is that even though they are different, all of them diverge on the quantum phase transition while \eqref{Rz0} never does so. This analysis shows that it could be fruitful to analyze the parameter dependent phase-space metric obtained via translations  of the phase space variables. We show in section \ref{seccion acoplados} that this metric is also useful to study entanglement in Gaussian states.

\section{Generalized harmonic oscillator with a linear term}

 In order to understand these concepts in a three-dimensional parameter space, let us now consider a more general Hamiltonian than \eqref{GHO}, given by
\begin{equation}\label{HWq OAGL}
H=\frac{1}{2}\left[X q^2+Y(qp+p q)+Z p^2\right]+W q,
\end{equation}
where our parameters are $\{\lambda_i\}= \{W,X,Y,Z\}$ with $i=1,2,3,4$. This system has been used to study the classical analog of the quantum geometric tensor \cite{Gonzalez}.

We can also define creation and annihilation operators for this extended Hamiltonian. Indeed, by performing the linear canonical transformation $\hat{q}=Z^{1/2}\hat{Q} - \frac{WZ}{\omega^2}$, $\hat{p}=Z^{-1/2}\hat{P}- Y(Z^{-1/2}\hat{Q}-\frac{W}{\omega^2})$, where $\omega=\left(X Z-Y^2\right)^{1 / 2}$ with the condition $X Z-Y^2>0$, the Hamiltonian \eqref{HWq OAGL} can be written  as
\begin{equation}
    \hat{H}=\frac{1}{2}  \left(\hat{P}^2 + \omega^2 \hat{Q}^2 - \frac{W^2 Z}{\omega^2}\right)  ,
\end{equation}
and then the creation and annihilation operators have the same form as in \eqref{creannhi ops OAG}. Notice that the energy is shifted by the factor $-\frac{W^2 Z}{2\omega^2}$ compared to the previous case.

Following the same procedure as before, the deformation functions are
\begin{equation} 
\mathcal{O}_W(t)= q, \quad 
	\mathcal{O}_X (t) = \frac{q^2}{2}, \quad \mathcal{O}_Y (t)=\frac{pq+ qp}{2}, \quad \mathcal{O}_Z(t)=\frac{p^2}{2}, 
\end{equation}

\begin{equation}
	\mathcal{O}_q(t)= Yp+ Xq + W, \quad 	\mathcal{O}_p (t)=Zp + Yq.
\end{equation}
and with them we get the quantum metric tensor
\begin{align}\label{qmtWq}
g_{i j}^{(n)}(x)=&\nonumber\frac{n+\frac{1}{2}}{  \omega^7}\left(\begin{array}{cccc}
Z \omega^4 & -W Z^2 \omega^2 & 2 W Y Z \omega^2 & -W Y^2 \omega^2 \\
-W Z^2 \omega^2 & W^2 Z^3 & -2 W^2 Y Z^2 & W^2 Y^2 Z \\
2 W Y Z \omega^2 & -2 W^2 Y Z^2 & W^2 Z\left(3 Y^2+X Z\right) & -W^2 Y\left(Y^2+X Z\right) \\
-W Y^2 \omega^2 & W^2 Y^2 Z & -W^2 Y\left(Y^2+X Z\right) & W^2 X Y^2
\end{array}\right)\\
&\nonumber+\frac{n^2+n+1}{32 \omega^4}\left(\begin{array}{cccc}
0 & 0 & 0 & 0 \\
0 & Z^2 & -2 Y Z & 2 Y^2-X Z \\
0 & -2 Y Z & 4 X Z & -2 X Y \\
0 & 2 Y^2-X Z & -2 X Y & X^2
\end{array}\right) \\
\end{align}\label{BerryCurvFull}
and the Berry Curvature
\begin{align} \label{BCWq}
F_{i j}^{(n)}(x) =&\frac{n+\frac{1}{2}}{4 \omega^3}\left(\begin{array}{cccc}
0 & 0 & 0 & 0 \\
0 & 0 & -Z & Y \\
0 & Z & 0 & -X \\
0 & -Y & X & 0
\end{array}\right) \nonumber\\
&+\frac{1}{  \omega^6}\left(\begin{array}{cccc}
0 & 0 & W Z \omega^2 & -W Y \omega^2 \\
0 & 0 & -W^2 Z^2 & W^2 Y Z \\
-W Z \omega^2 & W^2 Z^2 & 0 & -W^2 Y^2 \\
W Y \omega^2 & -W^2 Y Z & W^2 Y^2 & 0
\end{array}\right),
\end{align}
which are exactly the same as those of obtained in \cite{Gonzalez} by following a different approach.

\begin{figure}[ht]
  \centering
  \captionsetup{justification=raggedright,
singlelinecheck=false
}
  \includegraphics[width=0.8\textwidth]{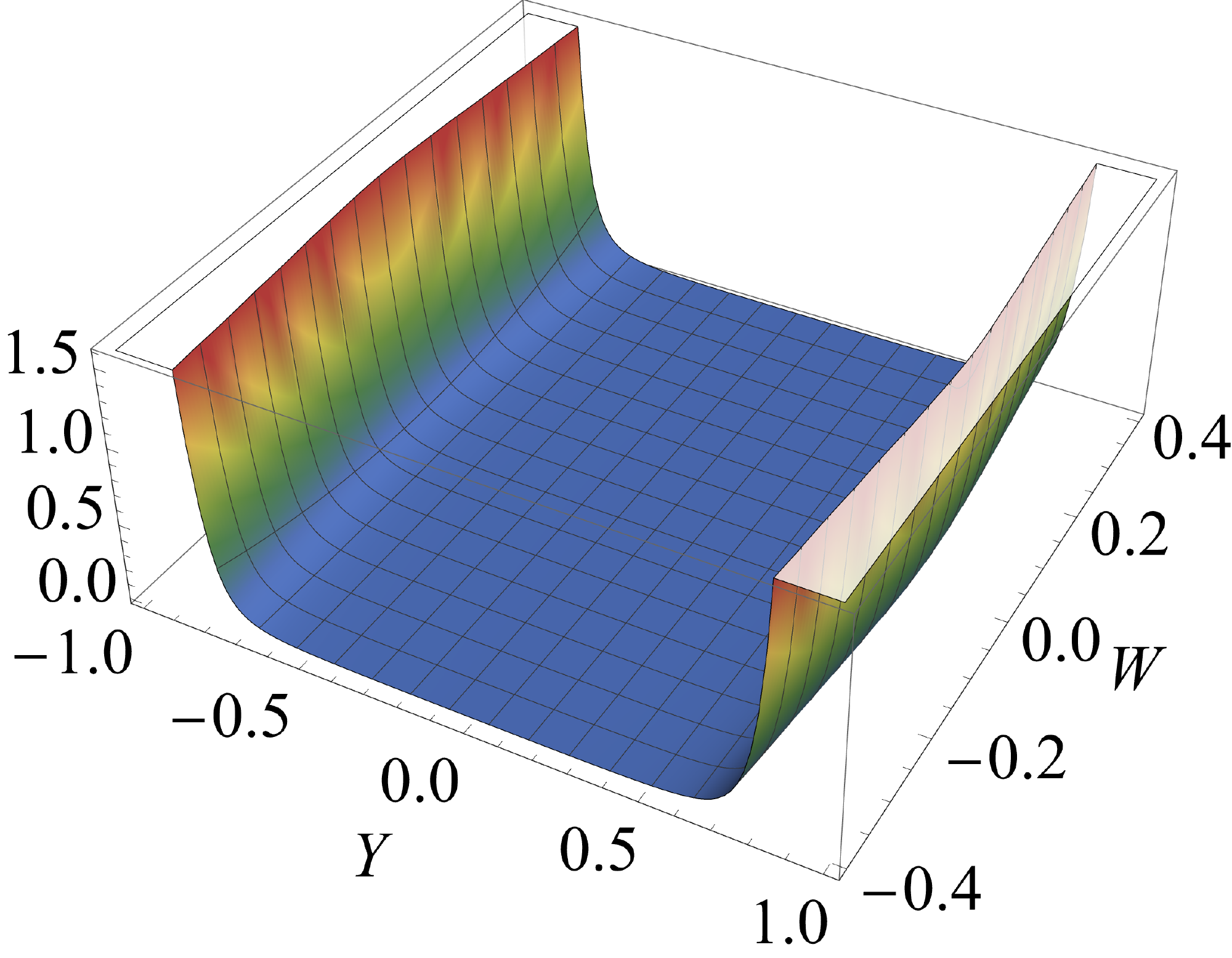}
  \captionof{figure}{Determinant \eqref{detQMT} of the parameter quantum metric tensor for $X=1$ and $n=0$.}
  \label{fig:OAGLdet3d}
\end{figure}

\begin{figure}[ht]
  \centering
  \captionsetup{justification=raggedright,
singlelinecheck=false
}
  \includegraphics[width=0.8\textwidth]{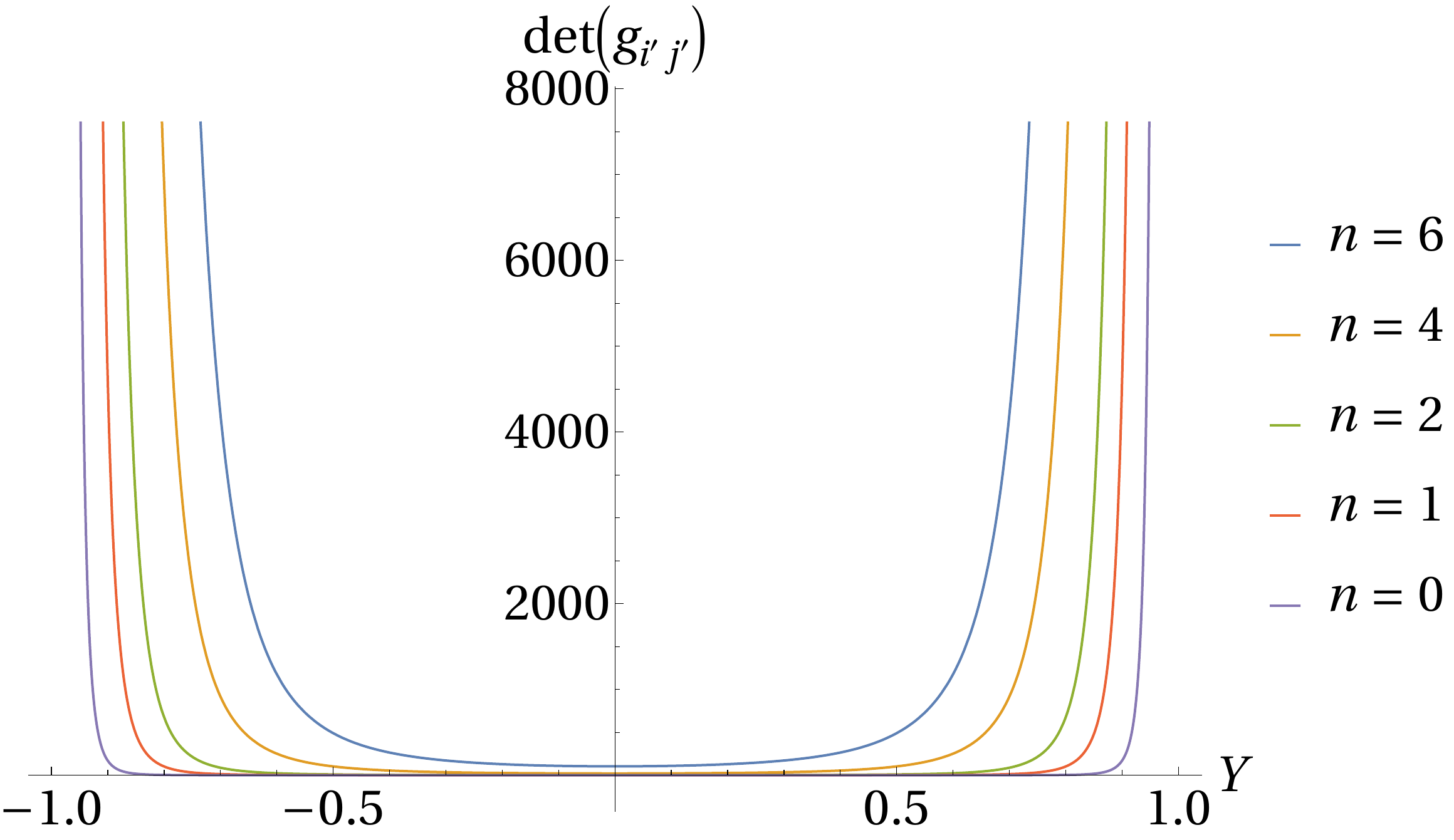}
  \captionof{figure}{Determinant \eqref{detQMT} of the parameter quantum metric tensor for different values of the quantum number $n$ when $X=1$ and $W=1$.}
  \label{fig:OAGLGrafdetYmultin}
\end{figure}

The metric $g_{i j}^{(n)}(x)$ has a vanishing determinant because it has rank three, meaning that there is a redundant parameter. In particular, if we set $Z=1$ and take $\left\{\lambda_{i'}\right\}=\{W, X, Y\}$ with $i'=1,2,3$ as the adiabatic parameters, the metric \eqref{qmtWq} reduces to
\begin{equation}\label{GHOL metric Z=1}
g^{(n)}_{i' j'}=\frac{a_n}{\omega^7}\left(\begin{array}{ccc}
\omega^4 & -W   \omega^2 & 2 W Y \omega^2 \\
-W   \omega^2 & W^2   & -2 W^2 Y   \\
2 W Y \omega^2 & -2 W^2 Y   & W^2\left(3 Y^2+X  \right)
\end{array}\right)+\frac{b_n}{32 \omega^4}\left(\begin{array}{ccc}
0 & 0 & 0 \\
0 & 1 & -2 Y \\
0 & -2 Y & 4 X
\end{array}\right),
\end{equation}
where
\begin{equation}
    a_n=  n+\frac{1}{2}  , \quad \text{and} \quad
b_n=n^2+n+1.
\end{equation}
The determinant of this metric is
\begin{equation}
\operatorname{det}\left[g_{i' j'}\right]=\frac{a_n b_n \left(8 a_n W^2 +b_n\omega^3 \right)}{256 \omega^{12}},
\end{equation}
or explicitly
\begin{equation}\label{detQMT}
   \operatorname{det}\left[g_{i' j'}\right]= \frac{\left(2 n^3+3 n^2+3 n+1\right) \left[\left(n^2+n+1\right)    \left(X-Y^2\right)^{3/2}+(8 n+4) W^2\right]}{512   \left(X-Y^2\right)^6},
\end{equation}
which is now different from zero. In Fig. \ref{fig:OAGLdet3d} we can see that the determinant has a divergence when  $\omega\to 0$, implying a phase transition. In Fig. \ref{fig:OAGLGrafdetYmultin} we notice that the divergent behavior of the determinant is noticeable sooner as the quantum number grows.


Furthermore, in this case, the Berry curvature \eqref{BCWq} reduces to
\begin{equation}
\begin{aligned}
F_{i' j'}^{(n)}(x) = & \frac{a_n}{8 \omega^3}\left(\begin{array}{ccc}
0 & 0 & 0 \\
0 & 0 & -1 \\
0 & 1 & 0
\end{array}\right) +\frac{1}{  \omega^6}\left(\begin{array}{ccc}
0 & 0 & W \omega^2 \\
0 & 0 & -W^2 \\
-W \omega^2 & W^2 & 0
\end{array}\right).
\end{aligned}
\end{equation}

\begin{figure}[h!]
  \centering
  \captionsetup{justification=raggedright,
singlelinecheck=false
}
  \includegraphics[width=0.8\textwidth]{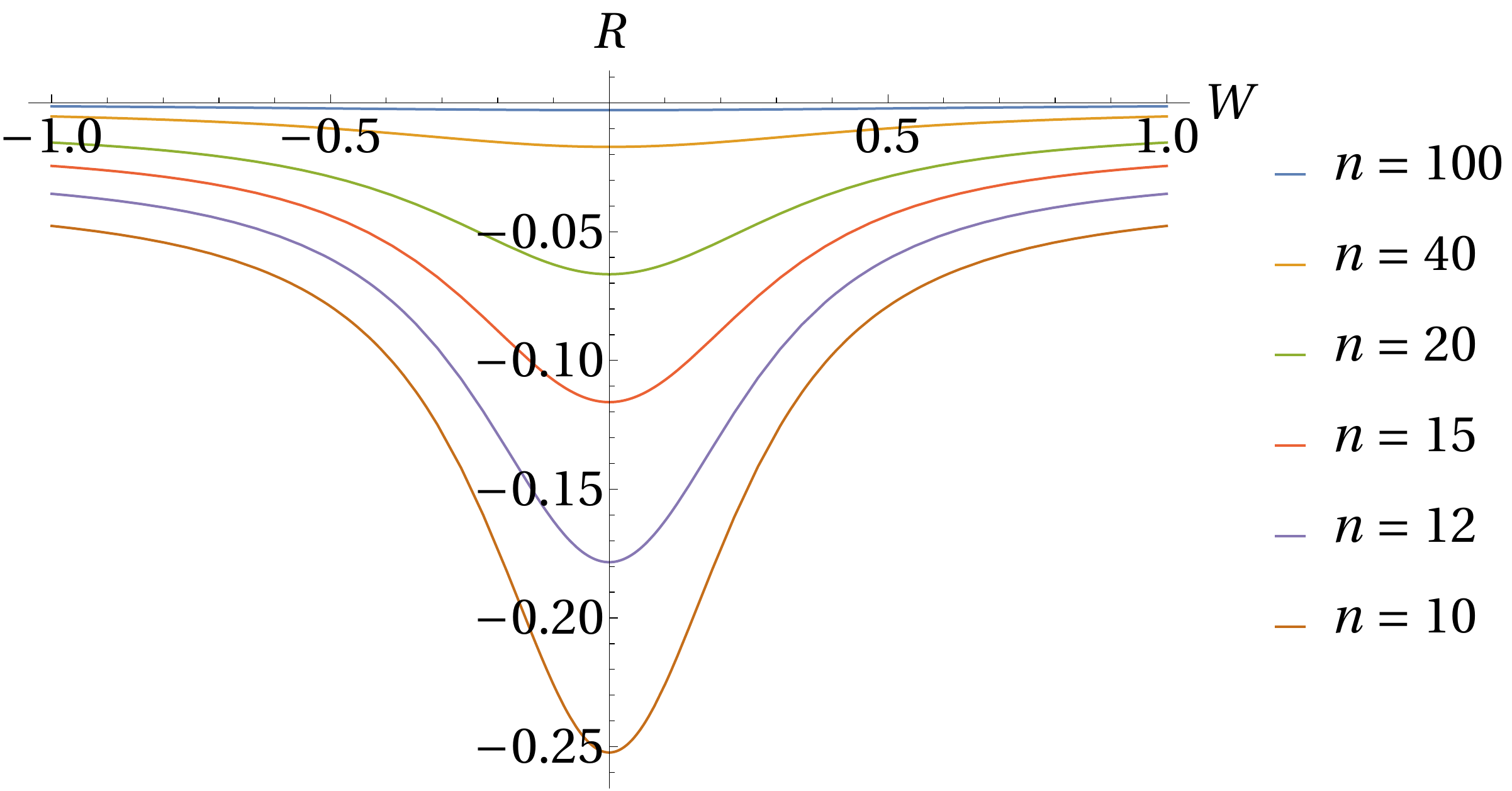}
  \captionof{figure}{Scalar curvature \eqref{scalarR} as a function of $W$ for different values of the quantum number $n$, fixing $X=1$ and $Y=0.9$. }
  \label{fig:OAGLescalarminfto}
\end{figure}

\begin{figure}[h!]
  \centering
  \captionsetup{justification=raggedright,
singlelinecheck=false
}
  \includegraphics[width=0.8\textwidth]{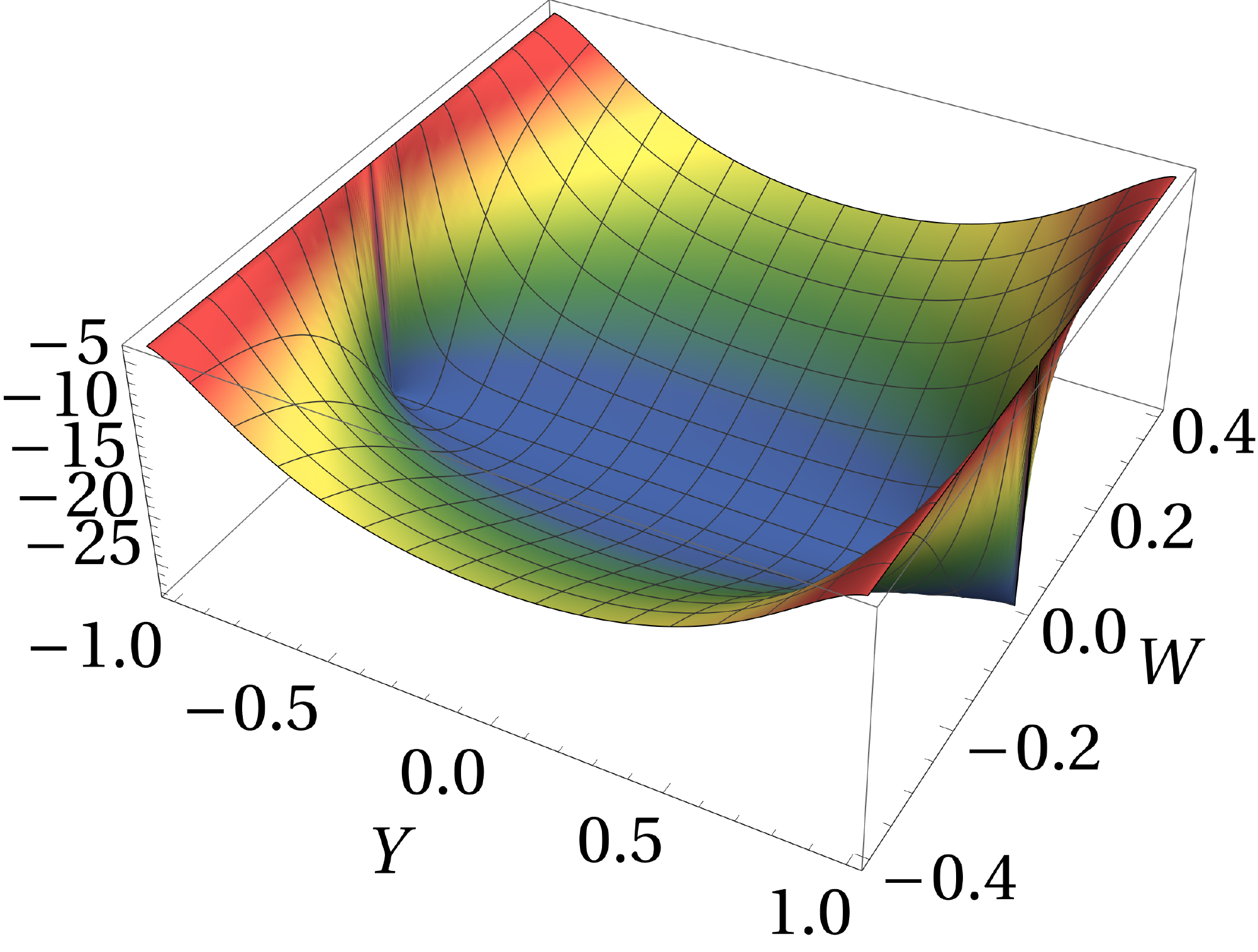}
  \captionof{figure}{Scalar curvature \eqref{scalarR} as a function of $Y$ and $W$   with $n=0$ and $X=1$.}
  \label{fig:OAGLescalar3DYW}
\end{figure}

Now that we have endowed our three-dimensional parameter space with the non-singular metric \eqref{GHOL metric Z=1}, we can study the corresponding scalar curvature. To this end, we begin by considering the inverse $g^{(n)\, i'j'}$ and computing the Christoffel symbols, which are in general defined by $\Gamma^i{}_{jk}=\frac{1}{2} g^{(n)\, il} (\partial_k g^{(n)}_{lj}+\partial_j g^{(n)}_{lk}-\partial_l g^{(n)}_{jk})$ \cite{Carroll} and turn out to be
\begin{align}
	\Gamma^{1}{}_{11} &=-\frac{8 a_n W}{b_n \omega^{3}} = -\Gamma^{2}{}_{21} ,& \quad
	\Gamma^{1}{}_{21} &=\frac{8 a_n W^2}{b_n \omega^{5}}-\frac{3}{4 \omega^2} ,& \quad
	\Gamma^{1}{}_{22} &= \frac{W}{2 \omega^4}-\frac{8 a_n W^3}{b_n \omega^{7}},
 \nonumber\\
	\Gamma^{1}{}_{31} &= \frac{3 Y}{2 \omega^2}-\frac{16 a_n W^2 Y}{b_n \omega^{5}},&
	\Gamma^{1}{}_{32} &= \frac{W Y \left(16 a_n W^2 -b_n \omega^3\right)}{b_n \omega^7},&		\Gamma^{2}{}_{11} &=-\frac{8 a_n}{b_n \omega},\nonumber\\	
	\Gamma^{2}{}_{22} &= -\frac{1}{\omega^2}-\frac{8 a_n W^2}{b_n \omega^{5}},& \Gamma^{2}{}_{31} &=-\frac{128 a_n^2 W^3 Y}{b_n \omega^2 \left(8 a_n W^2 \omega +b_n \omega^4\right)},&
	\Gamma^{3}{}_{31} &= \frac{8 a_n W }{8 a_n W^2 +b_n \omega^3},\nonumber\\
	\Gamma^{3}{}_{32} &=\frac{-20 a_n W^2-b_n \omega^{3}}{2 \omega^2 \left(8 a_n  W^2 + b_n \omega^3\right)},&
	\Gamma^{3}{}_{33} &=\frac{2 Y \left(20 a_n W^2+b_n \omega^{3}\right)}{\omega^2 \left(8 a_n W^2  +b_n \omega^3\right)},&
\nonumber
\end{align}
\begin{equation*}
    \Gamma^{2}{}_{32} =\frac{Y \left(128 a_n^2 W^4 \omega + 12 a_n b_n W^2 \omega^4+b_n^2 \omega^{7}\right)}{b_n \omega^{5} \left(8 a_n W^2 \omega+b_n \omega^4\right)}
\end{equation*}
 \begin{equation*}
     \Gamma^{2}{}_{33} = \frac{8 a_n W^2 \left(8 a_n W^2 \left(X-5 Y^2\right) \omega+b_n \omega^4\left(X+Y^2\right)\right)}{b_n \omega^{5} \left(8 a_n W^2 \omega+b_n \omega^4\right)},
 \end{equation*}
\begin{equation*}
    \Gamma^{1}{}_{33} = \frac{8 a W^3 \left(X-5 Y^2\right)}{b \omega^{7}}+\frac{W \left(X+Y^2\right)}{\omega^4}, \nonumber
\end{equation*}
These Christoffel symbols can also allow us to get geometric information of the parameter space through the geodesic equations \cite{Sarkar1}.
Using these results, the Ricci tensor entries $R_{i'j'}$ \cite{Carroll} are 
\begin{align}
	R_{11} &= \frac{2 a_n  \left(8 a_n W^2-3 b_n \omega^{3}\right)}{\left(8 a_n W^2  + b_n \omega^3\right)^2},& \quad
	R_{21} &=\frac{8 a_n W \left(2 a_n W^2 +b_n \omega^3\right)}{ \left(8 a_n W^2 \omega +b_n \omega^4\right)^2} ,\nonumber\\
	R_{22} &=\frac{-896 a_n^2 W^4 \omega -224 a_n b_N W^2 \omega^4-5 b_n^2 \omega^{7}}{16 \omega^{5} \left(8 a_n W^2  +b_n \omega^3\right)^2},&
	R_{31} &= -\frac{16 a_n W Y \left(2 a_n W^2  +b_n \omega^3\right)}{\omega^2 \left(8 a_n W^2  +b_n \omega^3\right)^2},\nonumber
\end{align}
\begin{equation*}
    R_{32} = -\frac{Y \left(-896 a_n^2 W^4 -224 a_n b_n W^2 \omega^3 -5 b_n^2\omega^{6}\right)}{8 \omega^{4} \left(8 a_n W^2  +b_n \omega^3\right)^2},
\end{equation*}
\begin{equation*}
    R_{33} =\frac{-64 a_n^2 W^4  \left(3 X+11 Y^2\right) -8 a_n b_n W^2 \left(9 X+19 Y^2\right) \omega^3 -b_n^2 \left(6 X-Y^2\right) \omega^{6}}{4 \omega^{4} \left(8 a_n W^2  +b_n \omega^3\right)^2}.
\end{equation*}
Using the inverse $g^{i'j'}$, the scalar curvature $R=g^{i'j'} R_{i'j'}$ is \footnote{Here we use the convention that repeated indices $i',j',\dots$ are summed from 1 to 3.}
\begin{equation}\label{scalarR}
    R=-\frac{4 \left(64 a_n^2 W^4+40 a_n b_n W^2 \omega^3 +7 b_n^2 \omega^{6}\right)}{b_n \left(8 a_n W^2  +b_n \omega^3\right)^2}.
\end{equation}
If $\omega\to 0$, then $R\to -4/b_n$, implying that this curvature does not diverge when the quantum phase transition occurs. In this case, the quantum phase transition is coordinate dependent. A remarkable feature of this scalar is that as the quantum number $n$ tends to infinity, it  becomes zero, i.e.
\begin{equation}
   \lim_{n\to \infty} R= 0,
\end{equation}
regardless of the value of $W$, as can see in Fig. \ref{fig:OAGLescalarminfto}.  Furthermore, in the case $W \to 0$ the scalar curvature \eqref{scalarR} becomes  $\lim_{W\to 0} R= -28/b_n$. As for the case of the ground state, $n=0$, it reduces to
\begin{equation}
  R=-\frac{4 \left(16 W^4+20 W^2 \omega ^3   +7 \omega ^6  \right)}{\left(4 W^2+\omega ^3  \right)^2}.
\end{equation}
In Fig. \ref{fig:OAGLescalar3DYW} we plot the scalar curvature as a function of $Y$ and $W$ setting $X=1$. Notice how its minimum $R=-28$ is obtained when $W=0$.

 The metric derived from the variation of the phase-space coordinates corresponds exactly with \eqref{5 TGC OAG pq}, meaning it is blind to the linear term of the Hamiltonian.

\section{Gaussian states}

The aim of the section is to obtain the generic form of the quantum geometric tensor for a system in a Gaussian state, whose wave function is
\begin{equation}\label{Wavefunction gauss}
   \psi(q,\lambda_1,\lambda_2)= \frac{1}{\sqrt{\sigma}\pi^ {1/4}} e^{-\frac{(q-\mu )^2}{2 \sigma ^2}},
\end{equation}
where $\sigma=\sigma(\lambda_1,\lambda_2)$, $\mu=\mu(\lambda_1,\lambda_2)$ are general functions of the adiabatic parameters $\{\lambda_i\}=\{\lambda_1,\lambda_2\}$  ($i=1,2$). Notice that given that this wave function is entirely real, then the Berry Curvature is automatically zero. In this case, using Provost's and Vallee's formulation \cite{Provost}, the quantum metric tensor is
\begin{equation}\label{metricgauss}
    g_{ij}=\frac{1}{2\sigma^2}\left(
\begin{array}{cc}
 (\partial_{\lambda_1} \sigma )^2+(\partial_{\lambda_1} \mu )^2 & \partial_{\lambda_2} \sigma \partial_{\lambda_1} \sigma +\partial_{\lambda_2} \mu  \partial_{\lambda_1} \mu  \\
 \partial_{\lambda_2} \sigma \partial_{\lambda_1} \sigma +\partial_{\lambda_2} \mu  \partial_{\lambda_1} \mu  & (\partial_{\lambda_2} \sigma)^2+(\partial_{\lambda_2} \mu) ^2 \\
\end{array}
\right),
\end{equation}
with determinant
\begin{equation}
    \operatorname{det}[g_{ij}]=\frac{\left(\partial_{\lambda_1} \sigma  \partial_{\lambda_2} \mu -\partial_{\lambda_2} \sigma \partial_{\lambda_1} \mu \right)^2}{4 \sigma^4}.
\end{equation}

Now, the non-zero Ricci tensor entries are
\begin{equation}
    R_{11}= -\frac{(\partial_{\lambda_1} \sigma )^2+(\partial_{\lambda_1} \mu )^2}{\sigma^2}
\end{equation}
\begin{equation}
    R_{22}=  -\frac{\partial_{\lambda_2} \sigma \partial_{\lambda_1} \sigma +\partial_{\lambda_2} \mu  \partial_{\lambda_1} \mu }{\sigma^2}
\end{equation}
\begin{equation}
    R_{22}=  -\frac{(\partial_{\lambda_2} \sigma)^2+(\partial_{\lambda_2} \mu )^2}{\sigma^2}
\end{equation}
giving a constant scalar curvature independent of any of the parameters
\begin{equation}\label{escalar -4}
    R=-4.
\end{equation}

To connect this section with the previous example, let us consider the wave function of the Hamiltonian \eqref{HWq OAGL}, which is
\begin{equation}\label{wavefunction GHOL}
\psi_n(q)=\left(\frac{\omega}{Z  }\right)^{1 / 4} \chi_n\left[\left(q+\frac{W Z}{\omega^2}\right) \sqrt{\frac{\omega}{Z  }}\right] \exp \left(-\frac{i Y q^2}{2 Z  }\right),
\end{equation}
where the functions $\chi(x)$ are given by
\begin{equation}
\chi_n(x)=\frac{1}{\sqrt{2^n \sqrt{\pi} n !}} e^{-x^2 / 2} H_n(x)
\end{equation}
with $H_n$ the Hermite polynomials.
Notice that \eqref{Wavefunction gauss} is a particular case of this wave function with $Y=0$, $Z=1$, $n=0$, $\sigma=X^{-1/4}$ and $\mu=W/X$. Using these identifications, \eqref{metricgauss} yields the metric \eqref{GHOL metric Z=1} after eliminating the corresponding rows and columns of the now constant parameters $Y$ and $Z$. Therefore, the scalar curvature that we get is \eqref{escalar -4}.

Up to this point we have only dealt with systems of one degree of freedom, so in the next section we study a system with two entangled particles.

\section{Symmetrically coupled  harmonic oscillators}\label{seccion acoplados}

Let us now consider a  system of two symmetrically coupled harmonic oscillators, which is quite interesting because it presents an interaction between their degrees of freedom. This system has many applications since it reveals entanglement between the oscillators \cite{ZeroModes} and, in the case of an infinite number of coupled oscillators, it generalizes to the most simple field theory \cite{Jefferson2017}. The Hamiltonian is
\begin{equation}\label{Hsym}
	H=\frac{1}{2}[p^2_1+p^2_2+k_0(q^2_1 + q^2_2)+k_1(q_1 - q_2)^2],
\end{equation}
where $\{\lambda_i\} =\{k_0, k_1\}$ $(i= 1, 2)$ are the adiabatic parameters. To simplify the Hamiltonian we transform to normal coordinates  using the canonical transformations $Q_1=\frac{1}{\sqrt{2}}(q_1+q_2)$, $Q_2=\frac{1}{\sqrt{2}}(q_1-q_2)$, $P_1=\frac{1}{\sqrt{2}}(p_1+p_2)$, and $P_2=\frac{1}{\sqrt{2}}(p_1-p_2)$, giving rise to a Hamiltonian corresponding to two uncoupled harmonic oscillators, namely $H=\frac{1}{2}(P_1^2+P_2^2+\omega_1^2 Q_1^2+\omega_2^2 Q_2^2)$ with frequencies
\begin{equation}
	\omega_1=\sqrt{k_0} \qquad {\rm and} \qquad\omega_2=\sqrt{k_0+2k_1},
\end{equation} 

Using \eqref{Hsym}, the deformation functions needed to compute the quantum geometric tensor (\ref{eq:defQGT}) are 
\begin{subequations}
    \begin{align}
        \mathcal{O}_{k_0}(t) &= \frac{\partial H}{\partial k_0} = \frac{1}{2}  (q_1^2 + q_2^2),\label{ok0} \\
        \mathcal{O}_{k_1}(t)&= \frac{\partial H}{\partial k_1} = \frac{1}{2}  (q^2_1 -q_1q_2 - q_2q_1 + q^2_2),\label{ok1}
    \end{align}
\end{subequations}
\begin{subequations}
    \begin{align}
        \mathcal{O}_{q_1}(t)&= \frac{\partial H}{\partial q_1} = \frac{1}{2} \left(\omega _2^2-\omega_1^2\right) (q_1-q_2)+\omega_1^2 q_1,\label{oq1}\\
        \mathcal{O}_{q_2}(t)&= \frac{\partial H}{\partial q_2}=\frac{1}{2} \left(\omega _2^2-\omega_1^2\right) (q_2-q_1)+\omega_1^2 q_2,\label{oq2} \\
        \mathcal{O}_{p_1} (t)&= \frac{\partial H}{\partial p_1}= p_1, \ \ \ \ \mathcal{O}_{p_2}(t) = \frac{\partial H}{\partial p_2}= p_2. \label{ops}
    \end{align}
\end{subequations}
Taking into account \eqref{ok0} and \eqref{ok1}, and following a procedure analogous to that used for the previous examples, the resulting quantum metric tensor for excited states is  
\begin{equation}\label{5 TGC OASA it}
	g_{ ij}^{(mn)} =\frac{1}{32}
	\left(
	\begin{array}{cc}
		\left(\frac{m^2+m+1}{\omega_1^4}+\frac{n^2+n+1}{\omega_2^4}\right) & \frac{2(n^2+n+1)}{ \omega_2^4} \\
		\frac{2(n^2+n+1)}{ \omega_2^4} & \frac{4(n^2+ n+1)}{ \omega_2^4} \\
	\end{array}
	\right),
\end{equation}
and can be simplified to
\begin{equation}\label{sco:metric}
	g^{(mn)}_{ ij}=(m^2+m+1)\frac{\partial_i \omega_1 \partial_j \omega_1}{8\omega_1^2} +(n^2+n+1) \frac{\partial_i \omega_2 \partial_j \omega_2}{8\omega_2^2},
\end{equation}
where $m$ and $n$ are non-negative integers. This metric is generalization to excited states of the metric for the ground state obtained in \cite{Alvarez2019}.  Notice that \eqref{5 TGC OASA it} has a nonvanishing determinant
\begin{equation}
	\det\left(g^{(mn)}_{ij}\right)=\frac{(m^2+m+1)(n^2+n+1)}{256\omega_{1}^{4}\omega_{2}^{4}}.
\end{equation}

Using the inverse $g^{(mn)\, ij}$ of \eqref{5 TGC OASA it}, we find that the only nonzero Christoffel symbols $\Gamma^i{}_{jk}$ in the parameter space are
\begin{align}\label{sco:Christoffel}
	\Gamma^{1}{}_{11}=-\frac{1}{k_0}, \qquad \Gamma^{2}{}_{11}=\frac{k_1}{k_0(k_0+2k_1)} , \nonumber \\  \Gamma^{2}{}_{12}=\Gamma^{2}{}_{21}=-\frac{1}{k_0+2k_1}, \qquad \Gamma^{2}{}_{22}=-\frac{2}{k_0+2k_1} . \nonumber
\end{align}
Then, as in the previous example, the resulting Christoffel symbols do not depend on the quantum numbers $m$ and $n$, in this way the Christoffel symbols and the geodesic equations are blind to the the energy level of the state. Now, plugging them into the expression of the Riemann curvature tensor $R^{i}{}_{jkl}=\partial_k \Gamma^i{}_{lj}-\partial_l \Gamma^i{}_{kj}+\Gamma^s{}_{lj}\Gamma^i{}_{ks}-\Gamma^s{}_{kj}\Gamma^i{}_{ls}$, we get
\begin{equation}
	R^{i}{}_{jkl}=0,
\end{equation}
which entails that the parameter space associated to the symmetric coupled harmonic oscillators is flat. This in turn means that we can introduce a new coordinate system $y=\{y^{i}\}=(u,v)$ so that $ \Gamma^i{}_{jk}=0$.  In such coordinates, the line element $ds^2=g^{(mn)}_{ij}d\lambda^i d\lambda^j$ is manifestly flat $ds^2=du^2+dv^2$ (we use the signature ``$+$'' since $\det(g^{(mn)}_{ij})>0$). By solving Beltrami's equation 
\begin{align}
	0=&\nonumber\frac{\partial}{\partial k_0} \left[\sqrt{\det(g^{(mn)}_{ij})} \left( g^{(mn)\, 11} \frac{\partial u}{\partial k_0} + g^{(mn)\, 12} \frac{\partial u}{\partial k_1}  \right)  \right] \\ &+ \frac{\partial}{\partial k_1} \left[\sqrt{\det(g^{(mn)}_{ij})} \left( g^{(mn)\, 12} \frac{\partial u}{\partial k_0} + g^{(mn)\, 22} \frac{\partial u}{\partial k_1}  \right)  \right],
\end{align}
and the equations
\begin{equation}
	\frac{\partial v}{\partial k_0} = -\sqrt{\det(g^{(mn)}_{ij})} \left( g^{(mn)\, 12} \frac{\partial u}{\partial k_0} + g^{(mn)\, 22} \frac{\partial u}{\partial k_1}  \right),
\end{equation}
\begin{equation}
     \frac{\partial v}{\partial k_1} = \sqrt{\det(g^{(mn)}_{ij})} \left( g^{(mn)\, 11} \frac{\partial u}{\partial k_0} + g^{(mn)\, 12} \frac{\partial u}{\partial k_1}  \right)
\end{equation}
the new coordinate system turns out to be
\begin{align}
	u(\lambda)&=\sqrt{\frac{(m^2+m+1)(n^2+n+1)}{128[(m^2+m+1)+(n^2+n+1)]}} \ln\left[ \det(g^{(mn)}_{ij}) \right], \\
	v(\lambda)&=\frac{ (m^2+m+1) \ln k+ (n^2+n+1) \ln k^{\prime} }{\sqrt{32[(m^2+m+1)+(n^2+n+1)]}}.
\end{align}

Now, we will focus on the phase space contribution to the generalized quantum geometric tensor, $G_{\alpha\beta}^{(n)}$. To calculate it, its useful to realize from the beginning that all the functions (\ref{oq1}, \ref{oq2}, \ref{ops}) are first order terms and the one-term expectation values of \eqref{eq:defQGT} vanish since the states are orthonormal. Then, only the factors $\langle\hat{{\cal O}}_{a}(t_{1})\hat{{\cal O}}_{b_n}(t_{2})\rangle_{n}$ contribute to the integral involved in \eqref{eq:defQGT}. In this way, for the phase-space part of the generalized quantum geometric tensor, with $\{r_\alpha\}=\{q_1,q_2, p_1, p_2\} $, we get 
\begin{align}\label{GQGTsym}
	G_{\alpha \beta}^{(mn)}=& \frac{1}{4 }\left(
	\begin{array}{cccc}
		c_m  \omega_1 + c_n  \omega_2  & c_m  \omega_1 - c_n  \omega_2 & 0 & 0 \\
		c_m  \omega_1 -c_n  \omega_2 & c_m  \omega_1 + c_n  \omega_2 & 0 & 0 \\
		0 & 0 & \frac{c_m}{ \omega_1 } + \frac{c_n }{ \omega_2  } & \frac{c_m  }{ \omega_1   } -\frac{c_n }{ \omega_2  } \\
		0 & 0 & \frac{c_m }{ \omega_1  } - \frac{c_n }{ \omega_2  } & \frac{c_m }{ \omega_1 } + \frac{c_n }{ \omega_2  } \\
	\end{array}
	\right) \nonumber
	\\&+
	\frac{i}{2 }\left(
	\begin{array}{cccc}
		0 & 0& 1 & 0 \\
		0 & 0& 0 & 1 \\
		1 & 0 & 0 & 0\\
		0 & 1 & 0 & 0 \\
	\end{array}
	\right).
\end{align}
where in this case $c_m =2m+1$ and $c_n =2n+1$.

\begin{figure}[h!]
    \centering
    \includegraphics[width=0.75\textwidth]{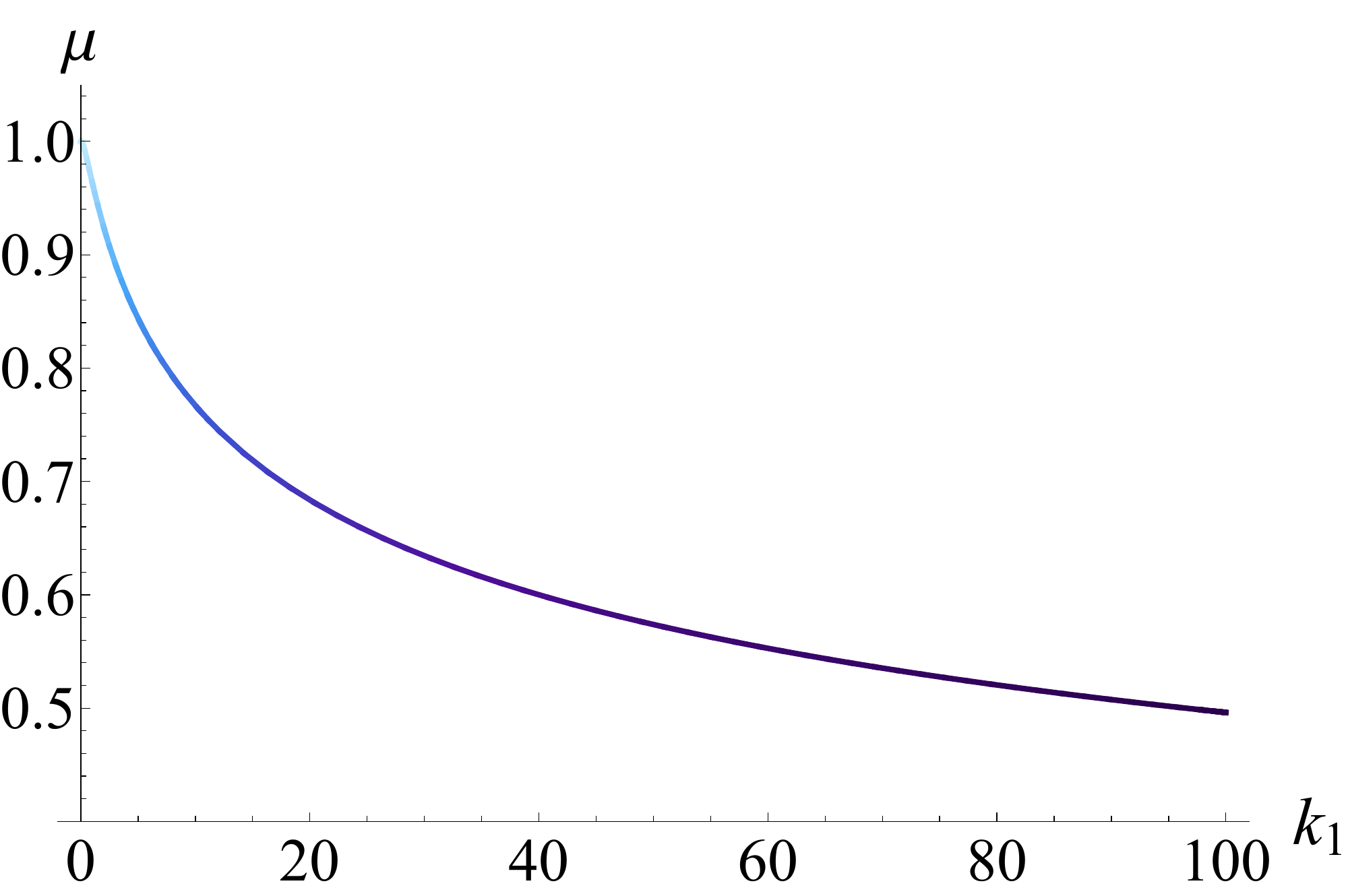}
    \caption{Purity \eqref{puritysym} of any of the oscillators as a function of $k_1$ while setting $k_0 = 1$.} 
    \label{purity2d}
\end{figure}

Taking the real part of \eqref{GQGTsym} and using the relations of \eqref{qcov qgt 1}, \eqref{qcov qgt 2}, and \eqref{qcov qgt 3}, we can get the quantum covariance matrix
\begin{align}
    \sigma^{(mn)}&=
\frac{1}{4}\begin{pmatrix}
 \frac{c_m }{ \omega_1  } + \frac{c_n }{ \omega_2  } & \frac{c_m }{ \omega_1 } - \frac{c_n }{ \omega_2  } & 0 & 0 \\
 \frac{c_m }{ \omega_1  } - \frac{c_n }{ \omega_2  } & \frac{c_m }{ \omega_1 } + \frac{c_n }{ \omega_2  } & 0 & 0 \\
0 & 0 & c_m  \omega_1 + c_n  \omega_2  & c_m  \omega_1 - c_n  \omega_2  \\
0 & 0 & c_m  \omega_1 - c_n  \omega_2  & c_m  \omega_1 + c_n  \omega_2  \end{pmatrix}.
\end{align}
To calculate both the purity and entropy of the reduced subsystems, we set $m=0$ and $n=0$ and consider the reduced quantum covariance matrices for each of our oscillators:
\begin{equation}\label{reducedsym}
    \sigma_1 = \sigma_2 =
\frac{1}{4}\begin{pmatrix}
\frac{1}{\omega_1}+\frac{1}{\omega_2} &   0 \\
0 & \omega_1 + \omega_2   
\end{pmatrix}.
\end{equation}
Notice that in this particular case these matrices are  the same for both oscillators given the symmetry between them in the Hamiltonian. Now, the determinant of \eqref{reducedsym} is
\begin{equation}
\operatorname{det}(\sigma_{1})=\operatorname{det}(\sigma_{2})=\frac{1}{4} \frac{\omega_{1}+\omega_{2}}{\sqrt{\omega_{1} \omega_{2}}},
\end{equation}
and hence the purity \eqref{purity gaussian} of the reduced subsystems yields
\begin{equation}\label{puritysym}
\mu = \mu(1)=\mu(2)=\frac{2 \sqrt{\omega_{1} \omega_{2}}}{\omega_{1}+\omega_{2}},
\end{equation}
In Fig. \ref{purity2d} we plot \eqref{puritysym} as a function of the coupling constant $k_1$. Notice that the purity decreases as the coupling constant increases and that the purity is 1 when the oscillators decouple i.e. when $k_1=0$ or $\omega_{1}=\omega_{2}$. On the other hand,  $\mu \to 0 $ as $k_0 \to 0$ or $k_1 \to \infty$.

\begin{figure}[h!]
    \centering
    \includegraphics[width=0.75\textwidth]{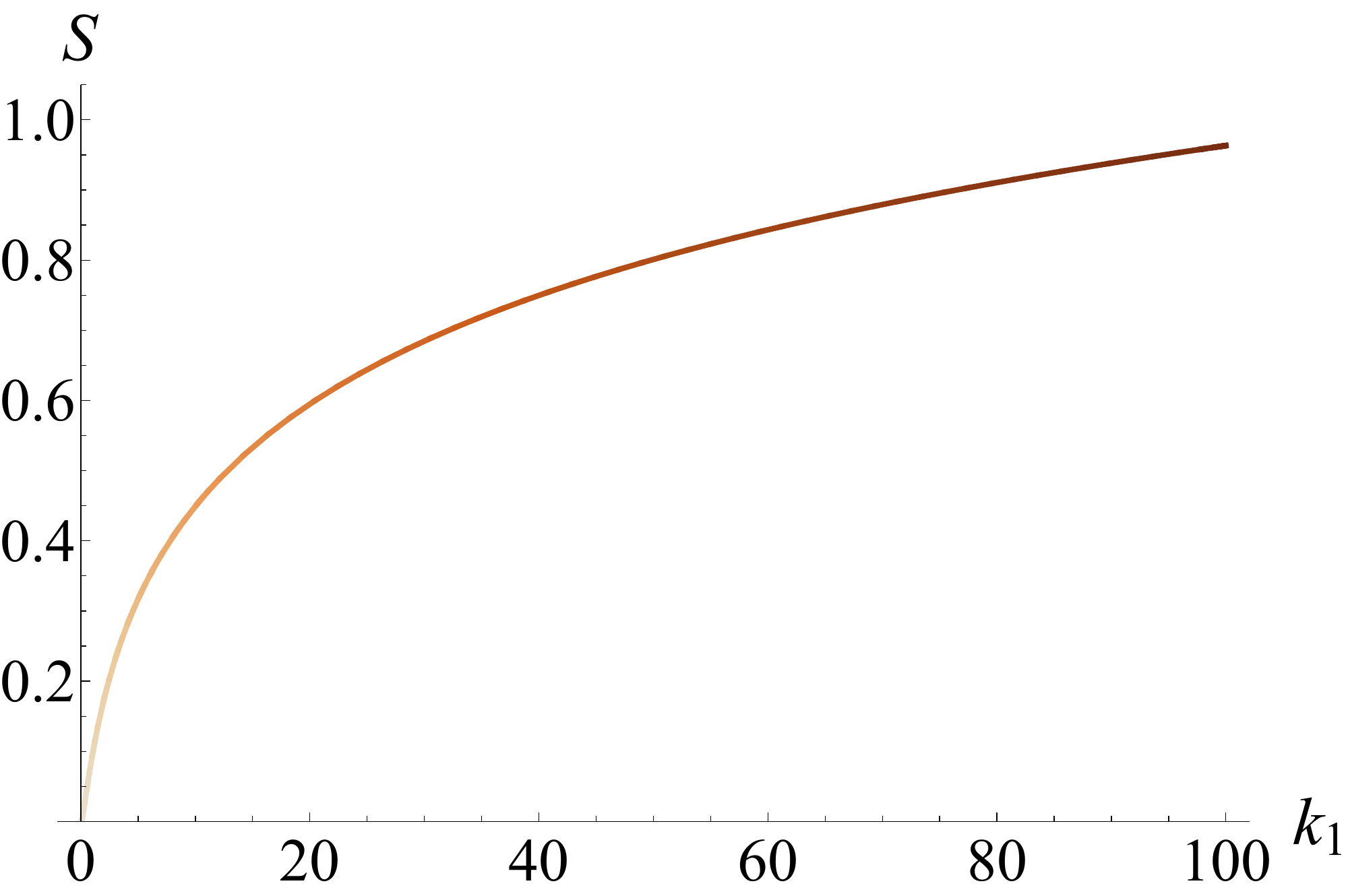}
    \caption{Entropy \eqref{entropysym} of any of the oscillators as a function of $k_1$ while setting $k_0 = 1$.}
    \label{entropy2d}
\end{figure}

For the von Neumann entropy, we calculate the symplectic eigenvalue of either $\sigma_1 $ or $\sigma_2$, which turns out to be
\begin{equation}
    \nu= \frac{\omega_1 + \omega_2}{4 \sqrt{\omega_1 \omega_2 }}.
\end{equation}
Using this result, the entropy \eqref{von Neumann OASA} leads to
\begin{align}\label{entropysym}
S=S_1\left(\nu\right)=S_2\left(\nu\right)= &\left(\left(\frac{\omega_1 + \omega_2}{4 \sqrt{\omega_1 \omega_2 }}\right)+\frac{1}{2}\right) \ln \left(\left(\frac{\omega_1 + \omega_2}{4 \sqrt{\omega_1 \omega_2 }}\right)+\frac{1}{2}\right) \nonumber
\\&-\left(\left(\frac{\omega_1 + \omega_2}{4 \sqrt{\omega_1 \omega_2 }}\right)-\frac{1}{2}\right) \ln \left(\left(\frac{\omega_1 + \omega_2}{4 \sqrt{\omega_1 \omega_2 }}\right)-\frac{1}{2}\right),
\end{align}
In Fig. \ref{entropy2d} we show the entropy as a function of $k_1$. We recognize that the entropy increases as the coupling constant $k_1$ strengthens, but also that as the self coupling constant $k_0$ decreases the entropy increases, i.e. $S \to \infty$ as $k_0 \to 0$. From this we notice that reducing the self coupling increases the entanglement.

With this we have seen that the generalized quantum geometric tensor indeed has all the information related to the quantum covariance matrix, allowing us to calculate the purity and von Neumann entropy of Gaussian states.

Now, to calculate the scalar curvature of the phase space contribution of the generalized quantum geometric tensor \eqref{GQGTsym} it must first be noticed that in this system we only have two parameters and a $4\times4$ matrix. Then instead we will analyze the geometry of the reduced matrices pertaining to each of the particles, since these can be related to the reduced quantum covariance matrices \eqref{reducedsym}. These reduced matrices are
\begin{equation}\label{OASApqreducida}
    g_{\alpha' \beta'}^{(mn)}= \frac{1}{4 }
	\begin{pmatrix}
		c_m  \omega_1 + c_n  \omega_2  & 0 \\
		0 & \frac{c_m }{ \omega_1 } + \frac{c_n }{ \omega_2  } \\
	\end{pmatrix},
\end{equation}
where $\alpha',\beta' = 1,2$. We will consider that \eqref{OASApqreducida} is a metric of parameter space with coordinates $\{k_0,k_1\}$. Taking into account this consideration, the scalar curvature is
\begin{align}
    R=&\frac{1}{2  \omega_1 ^4  \omega_2 ^4 ( c_m   \omega_2 + c_n   \omega_1 )^2 ( c_m   \omega_1 + c_n   \omega_2 )^2}\bigg[-6  c_m ^3  \omega_1   \omega_2 ^6-6  c_n ^3  \omega_1 ^6  \omega_2 \nonumber\\ \nonumber& +  c_m ^2  c_n   \omega_2  \left(-6  \omega_1 ^6+ \omega_1 ^4  \omega_2 ^2-8  \omega_1 ^2  \omega_2 ^4-5  \omega_2 ^6\right)\\&\nonumber+ c_m   c_n ^2  \omega_1  \left(-5  \omega_1 ^6-8  \omega_1 ^4  \omega_2 ^2+ \omega_1 ^2  \omega_2 ^4-6  \omega_2 ^6\right) \\ & + 4  c_n   \omega_1 ^5  \omega_2 ^2 \left(2  \omega_1   \omega_2  \left( c_m ^2+ c_n ^2\right)+ c_m   c_n   \omega_1 ^2+3  c_m   c_n   \omega_2 ^2\right)\bigg].
\end{align}

\begin{figure}[ht]
  \centering
  \captionsetup{justification=raggedright,
singlelinecheck=false
}
  \includegraphics[width=0.8\textwidth]{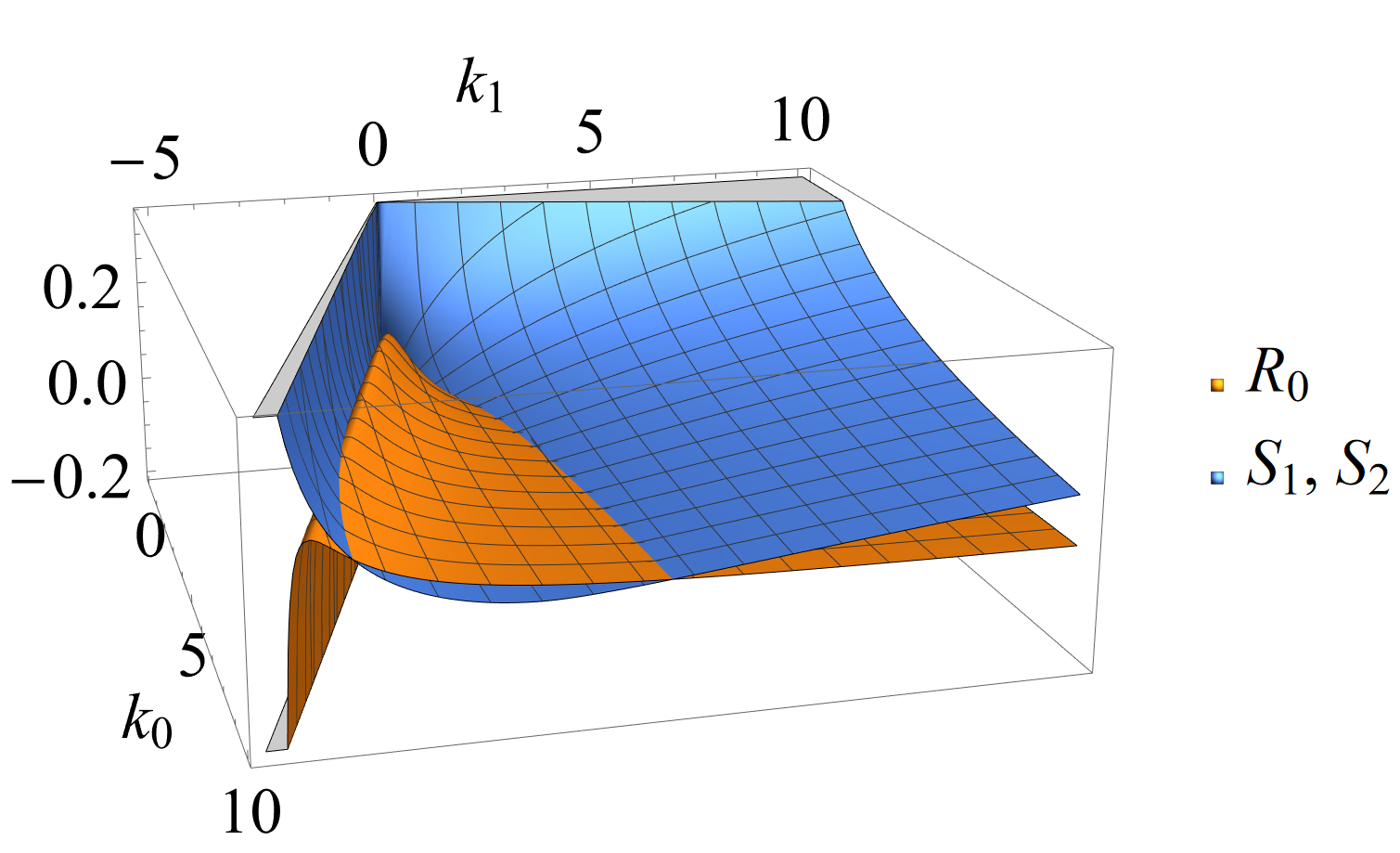}
  \captionof{figure}{Scalar curvature \eqref{R0 oasa pq} corresponding to the phase-space variation and the von Neumann entropy \eqref{von Neumann OASA} for the ground state of the symmetrically coupled oscillators.}
  \label{fig:OASAR0vsS}
\end{figure}

That comparatively to $\mu(1)=\mu(2)$, and $S_1=S_2$ behaves as

\begin{equation}
    \lim_{k_0 \to \infty} R = 0 \qquad \lim_{k_0 \to \infty} \mu(1) = 1 \qquad \lim_{k_0 \to \infty} S_1 = 0 
\end{equation}
\begin{equation}
    \lim_{k_0 \to 0} R= -\infty \qquad \lim_{k_0 \to 0} \mu(1)= 0 \qquad  \lim_{k_0 \to 0} S_1= \infty
\end{equation}
\begin{equation}
    \lim_{k_1 \to \infty} R = 0 \qquad \lim_{k_1 \to \infty} \mu(1) = 0 \qquad \lim_{k_1 \to \infty} S_1 = \infty 
\end{equation}
\begin{equation}
    \lim_{k_1 \to 0} R= \frac{4 c_n k_0-3 (c_m+c_n)}{k_0^{5/2} (c_m+c_n)^2} \qquad \lim_{k_1 \to 0} \mu(1)= 1 \qquad  \lim_{k_1 \to 0} S_1= 0
\end{equation}
and for the ground state $m=n=0$ reduces to
\begin{equation}\label{R0 oasa pq}
    R_0= \frac{2  \omega_1  ( \omega_1 +3  \omega_2 )}{ \omega_2 ^2 ( \omega_1 + \omega_2 )^3}-\frac{\left( \omega_1 ^2+ \omega_1   \omega_2 + \omega_2 ^2\right) \left(5  \omega_1 ^2-8  \omega_1   \omega_2 +5  \omega_2 ^2\right)}{2  \omega_1 ^4  \omega_2 ^4 ( \omega_1 + \omega_2 )},
\end{equation}
whose behavior in comparison with the von Neumann entropy \eqref{von Neumann OASA} is plotted in Fig.~\ref{fig:OASAR0vsS}.

\section{Linearly coupled harmonic oscillators }

We now want to illustrate our path integral approach on a more general system of two coupled harmonic oscillators described by the Hamiltonian
\begin{equation}\label{Hlin}
	H=\frac{1}{2}(p_{1}^{2}+p_{2}^{2}+ Aq_{1}^{2}+Bq_{2}^{2}+Cq_{1}q_{2}),
\end{equation}
where $\{\lambda_i\} = \{A, B, C\}$ are the adiabatic parameters, we restrict ourselves to the region where $A \neq B$, and $4 A B-C \geqslant 0$. Notice that \eqref{Hsym} is not a particular case of this system, and thus the results of the preceding section cannot be obtained from the results we present here. The system under consideration has proven to be useful for studying quantum entanglement \cite{Makarov}. In particular, it has previously been demonstrated to exhibit significant quantum entanglement for certain parameter values \cite{Makarov2020}. Furthermore, this system has been utilized in \cite{Diego2} to investigate the classical counterpart of the quantum geometric tensor.

The deformation functions needed for this system are
\begin{subequations}
    \begin{align}
        \mathcal{O}_A(t) &= \frac{\partial H}{\partial A} = \frac{1}{2}  q_1^2, \\
        \mathcal{O}_{B}(t)&= \frac{\partial H}{\partial B} = \frac{1}{2}  q^2_2, \\
        \mathcal{O}_C(t) &= \frac{\partial H}{\partial C} = \frac{1}{2}  q_1q_2,
    \end{align}
\end{subequations}
\begin{subequations}
	\begin{align}
	    \mathcal{O}_{q_1}(t) = \frac{\partial H}{\partial q_1} &= A q_1 +\frac{1}{2} C q_2, \\
     \mathcal{O}_{q_2}(t)= \frac{\partial H}{\partial q_2}&=B q_2 +\frac{1}{2} C q_1, \\
     \mathcal{O}_{p_1}(t) = \frac{\partial H}{\partial p_1}= p_1, & \ \ \ \ \mathcal{O}_{p_2} (t)= \frac{\partial H}{\partial p_2}= p_2.
	\end{align}
\end{subequations}

To diagonalize the Hamiltonian, we perform the canonical transformation
\begin{equation}
Q=\mathcal{R} q, \quad P=\mathcal{R}  p,
\end{equation}
where
\begin{equation}
Q=\left(\begin{array}{l}
Q_1 \\
Q_2
\end{array}\right), \quad q=\left(\begin{array}{l}
q_1 \\
q_2
\end{array}\right), \quad P=\left(\begin{array}{l}
P_1 \\
P_2
\end{array}\right), \quad p=\left(\begin{array}{l}
p_1 \\
p_2
\end{array}\right),
\end{equation}
and
\begin{equation}
\mathcal{R} =\left(\begin{array}{cc}
\cos \zeta & -\sin \zeta \\
\sin \zeta & \cos \zeta
\end{array}\right) .
\end{equation}
Here, the angle $\zeta$ is such that $\tan\zeta = \frac{\epsilon}{|\epsilon|} \sqrt{\epsilon ^2+1}-\epsilon$ with $\epsilon = (B-A)/C$, such that $\zeta \in(-\pi / 4, \pi / 4)$. Notice that the transformation matrix $\mathcal{R}$ depends on the system's parameters. In terms of the new variables, the Hamiltonian \eqref{Hlin} reads
$
H=\frac{1}{2}\left(P_1^2+P_2^2+\omega_1^2 Q_1^2+\omega_2^2 Q_2^2\right),
$
where $\omega_1$ and $\omega_2$ are the normal frequencies
\begin{equation}
\omega_1=\sqrt{A-\frac{C}{2} \tan \zeta}, \quad \omega_2=\sqrt{B+\frac{C}{2} \tan \zeta} .
\end{equation}

Following the same procedure as before, the quantum metric tensor for excited states is
\begin{align}\label{lco:qmetric}
	g_{kl}^{(mn)}=& \nonumber\frac{m^{2}+m+1}{32\omega_{1}^{4}}M_{kl}+\frac{n^{2}+n+1}{32\omega_{2}^{4}}N_{kl} \\&+\frac{L_{kl}}{4\left(\omega_{2}^{2}-\omega_{1}^{2}\right)^{2}}\left[\left(\frac{\omega_{1}}{\omega_{2}}+\frac{\omega_{2}}{\omega_{1}}\right)\left(m+\frac{1}{2}\right)\left(n+\frac{1}{2}\right)-\frac{1}{2}\right],
\end{align}
where
\begin{align}
	M_{kl} & =\frac{1}{4}\begin{pmatrix}(1+\eta)^{2} & \varphi^{2} & -(1+\eta)\varphi\\
		\varphi^{2} & (1-\eta)^{2} & -(1-\eta)\varphi\\
		-(1+\eta)\varphi & -(1-\eta)\varphi & \varphi^{2}
	\end{pmatrix},\nonumber \\
	N_{kl} & =\frac{1}{4}\begin{pmatrix}(1-\eta)^{2} & \varphi^{2} & (1-\eta)\varphi\\
		\varphi^{2} & (1+\eta)^{2} & (1+\eta)\varphi\\
		(1-\eta)\varphi & (1+\eta)\varphi & \varphi^{2}
	\end{pmatrix},\nonumber \\
	L_{kl} & =\begin{pmatrix}\varphi^{2} & -\varphi^{2} & \eta\varphi\\
		-\varphi^{2} & \varphi^{2} & -\eta\varphi\\
		\eta\varphi & -\eta\varphi & \eta^{2}
	\end{pmatrix},
\end{align}
and we have defined $\eta\equiv\cos2\zeta=\frac{\epsilon}{\sqrt{\epsilon^{2}+1}}$
and $\varphi\equiv\sin2\zeta=\frac{1}{\sqrt{\epsilon^{2}+1}}$. It can be simplified to
\begin{align}\label{qgt OALA}
	g^{(mn)}_{ ij}=&(m^2+m+1)\frac{\partial_i \omega_1 \partial_j \omega_1}{8\omega_1^2} +(n^2+n+1) \frac{\partial_i \omega_2 \partial_j \omega_2}{8\omega_2^2}\nonumber\\& +\partial_i \zeta \partial_j \zeta\left[\left(\frac{\omega_1}{\omega_2}+\frac{\omega_2}{\omega_1}\right)\left(m+\frac{1}{2}\right)\left(n+\frac{1}{2}\right)-\frac{1}{2}\right].
\end{align}
This is a generalization of the result of \cite{Alvarez2019}.
The determinant of \eqref{lco:qmetric} is
\begin{equation} \label{det G par OALA}
	\det \left(g_{kl}^{(mn)}\right)=\frac{\left(m^{2}+m+1\right)\left(n^{2}+n+1\right)\left[\left(m+\frac{1}{2}\right)\left(n+\frac{1}{2}\right)\left(\omega_{1}^{2}+\omega_{2}^{2}\right)-\frac{\omega_{1}\omega_{2}}{2}\right]}{4096\,\omega_{1}^{5}\omega_{2}^{5}\left(\omega_{2}^{2}-\omega_{1}^{2}\right)^{2}}.
\end{equation}
Notice how a divergence occurs when $\omega_1=0$ or  $\omega_2 = 0$. The case when $\omega_1 = \omega_2$ has already been excluded since we would need $A=B$.

From this point onward we will assume that $\frac{\epsilon}{|\epsilon|}=+1$ in order to simplify the calculations. For the ground state, the nonzero Christoffel symbols constructed out of the metric tensor (\ref{lco:qmetric}) are 
\begin{align}
	\Gamma^{1}{}_{11} &= -\frac{(2 B+E)^2}{E^2 F},& \qquad
	\Gamma^{1}{}_{21} &=\Gamma^{2}{}_{21}= -\frac{C^2}{2 E^2 F} ,& \qquad
	\Gamma^{1}{}_{31} &= \frac{C (3 B+2 E)}{2 E^2 F},\nonumber\\
	\Gamma^{1}{}_{32} &= \frac{A C}{2 E^2 F},&
	\Gamma^{1}{}_{33} &= -\frac{A (2 B+E)}{E^2 F},&	
	\Gamma^{2}{}_{22} &= -\frac{(2 A+E)^2}{E^2 F},\nonumber\\	
	\Gamma^{2}{}_{31} &= \frac{B C}{2 E^2 F},&
	\Gamma^{2}{}_{32} &= \frac{C (3 A+2 E)}{2 E^2 F},&
	\Gamma^{2}{}_{33} &= -\frac{B (2 A+E)}{E^2 F},\nonumber\\
	\Gamma^{3}{}_{11} &=\frac{2 B C}{E^2 F} ,&
	\Gamma^{3}{}_{21} &= \frac{C (A+B+2 E)}{E^2 F} ,&
	\Gamma^{3}{}_{22} &= \frac{2 A C}{E^2 F},\nonumber\\
	\Gamma^{3}{}_{31} &= -\frac{B (3 A+B+2 E)}{E^2 F},&
	\Gamma^{3}{}_{32} &=-\frac{A (A+3 B+2 E)}{E^2 F} ,&
	\Gamma^{3}{}_{33} &=  \frac{C}{E^2},
\end{align}
where $E:=\sqrt{4 A B-C^2}$ and $F:=A+B+E$. By using these expressions, we compute the Riemann curvature tensor $R^{i}{}_{jkl}$, however, the result is long and uninstructive, and so we prefer not to write it down. Despite this, we find that the scalar curvature  $ R^{(00)} =g^{(00)\, ij}R^{k}{}_{ikj}$  has the very simple form
\begin{equation}\label{menos8}
	R^{(00)}=-8.
\end{equation}
Thus, the associated parameter space has constant negative scalar curvature. When one of the oscillators can be in any excited state, the scalar curvature is
\begin{align}
   R^{(0n)} =& -\frac{1}{\left(n^2+n+1\right) \left(2 n (A+B)+A+B -E\right)^3} \nonumber \\ & \bigg[4 (2 n+1) \left(n^2+n+2\right) \left(A^3 (2 n+1)^2+A^2 \left(B (28 n (n+1)+15)-3 (2 n+1) E\right) \right. \nonumber \\ &-A \left(10 B (2 n+1) E-\left(B^2 (28 n (n+1)+15)\right)+C^2 (4 n (n+1)+3)\right) \nonumber \\ &\left. -3 B^2 (2 n+1) E+C^2 (2 n+1) E +B^3 (2 n+1)^2-B C^2 (4 n (n+1)+3)\right) \bigg],
\end{align}
which when $n\to 0$ yields \eqref{menos8}. This curvature also does not exhibit any divergences for real values of its parameters.

The phase space contribution of the generalized quantum geometric tensor for this system is
\begin{equation}\label{Oalapqs}
  g^{(mn)}_{\alpha\beta} = \begin{pmatrix}
g_{q_1 q_2} & 0_{2\times2} \\ 0_{2\times2} & g_{p_1 p_2}
\end{pmatrix} 
\end{equation}
with
\begin{equation}
g_{q_1 q_2} =  \left(\begin{array}{ll}
c_m \omega_1 \cos ^2 \zeta+ c_n \omega_2 \sin ^2 \zeta & \left(c_n \omega_2- c_m\omega_1\right) \sin \zeta \cos \zeta \\
\left(c_n \omega_2- c_m \omega_1\right) \sin \zeta \cos \zeta & c_m  \omega_1 \sin ^2 \zeta+ c_n \omega_2 \cos ^2 \zeta
\end{array}\right)
\end{equation}
\begin{equation}
g_{p_1 p_2}= \left(\begin{array}{cc}
\frac{c_m \cos ^2 \zeta}{\omega_1}+\frac{c_n \sin ^2 \zeta}{\omega_2} & \left(\frac{c_n}{\omega_2}-\frac{c_m}{\omega_1}\right) \sin \zeta \cos \zeta \\
\left(\frac{c_n}{\omega_2}-\frac{c_m}{\omega_1}\right) \sin \zeta \cos \zeta & \frac{c_m \sin ^2 \zeta}{\omega_1}+\frac{c_n \cos ^2 \zeta}{\omega_2}
\end{array}\right)
\end{equation}
where $c_m = \left(m + \frac{1}{2}\right)$ and $c_n = \left(n + \frac{1}{2}\right)$. Now in order to construct the scalar curvature, we reduce the metric \eqref{Oalapqs} to only consider the first particle, getting
\begin{equation}\label{OALApqreducida1}
g_{\alpha' \beta'}^{(mn)}=\left(\begin{array}{cc}
\frac{c_m \cos ^2 \zeta}{\omega_1}+\frac{c_n \sin ^2 \zeta}{\omega_2} & 0\\
0 & c_m \omega_1 \cos ^2 \zeta+ c_n \omega_2 \sin ^2 \zeta
\end{array}\right),
\end{equation}
with $\alpha',\beta' = 1,2$, the determinant of \eqref{OALApqreducida1} is
\begin{equation}
    \det(g_{\alpha' \beta'}^{(mn)}) = \frac{\left(c_m  \omega_2 \cos ^2\zeta +c_n  \omega_1 \sin ^2\zeta \right) \left(c_m  \omega_1 \cos ^2\zeta +c_n  \omega_2 \sin ^2\zeta \right)}{\omega_1 \omega_2}.
\end{equation}
Notice that once again it carries the divergence present in \eqref{det G par OALA}.
We will consider that \eqref{OALApqreducida1} is a metric of parameter space, where we can consider any of the possible combinations of parameters as coordinates, i.e. $\{A,B\}$, $\{B,C\}$ or $\{A,C\}$. For the rest of this section we will use $\{B,C\}$ as coordinates, although we observe that $\{A,C\}$ behaves similarly. We were keen to select the parameter $C$ as coordinate since it is the coupling parameter and thus carries most of the information regarding entanglement.

\begin{figure}[ht]
  \centering
  \captionsetup{justification=raggedright,
singlelinecheck=false
}
\includegraphics[width=0.8\textwidth]{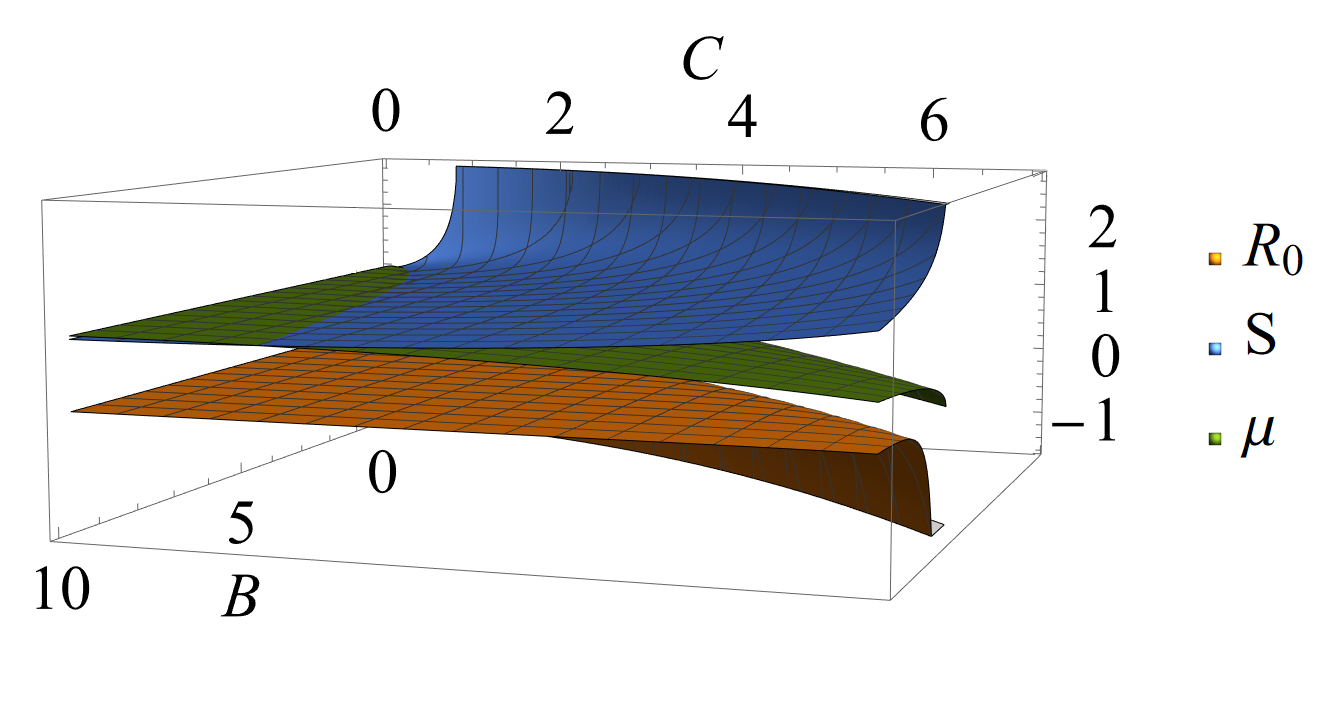}
  \captionof{figure}{Scalar curvature $R_0$  corresponding to \eqref{OALApqreducida1}, purity \eqref{purityOALA} and the von Neumann entropy \eqref{entropyOALA} for the ground state of the linearly coupled oscillators.}
  \label{fig:OALAR0vsSvsmu}
\end{figure}

Even thought is possible to calculate analytically the scalar curvature using, due to the definition of $\zeta$ it is too cumbersome to write explicitly. For the ground state, the purity is
\begin{equation}\label{purityOALA}
    \mu =\sqrt{\frac{4 A B-C^2}{4 A B}},
\end{equation}
and the von Neumann entropy
\begin{align}\label{entropyOALA}
    S=&\left(\sqrt{\frac{4 A B}{4 A B-C^2}}+\frac{1}{2}\right) \log \left(\sqrt{\frac{4 A B}{4 A B-C^2}}+\frac{1}{2}\right)\nonumber\\&-\left(\sqrt{\frac{4 A B}{4 A B-C^2}}-\frac{1}{2}\right) \log \left(\sqrt{\frac{4 A B}{4 A B-C^2}}-\frac{1}{2}\right).
\end{align}
In Fig. \ref{fig:OALAR0vsSvsmu}, we display these two quantities and the scalar curvature for the ground state using $\{B,C\}$. We observe that the scalar curvature for \eqref{OALApqreducida1} has a similar behavior compared with the purity. However, when the purity reaches zero, the scalar curvature diverges to negative values, indicating that it can recognize the maximum quantum entanglement as a phase transition.

\section{Conclusions}

In this paper, we accomplished two critical goals from a theoretical perspective. First, we provide a formulation of the quantum geometric tensor in the path integral formalism that is able to handle both the system's ground state and any excited state. This formalism enables us to detect excited state quantum phase transitions (ESQPT) by means of the divergent behavior of the quantum geometric tensor. We also show that the path integral formulation of the quantum geometric tensor can be extended to include translational variations of the phase-space coordinates in Hamiltonian formalism. This provides an alternative approach to calculate the quantum covariance matrix, and consequently an attractive method to obtain the purity and von Neumann entropy for a Gaussian state. Second, we prove the equivalence between the path integral formalism for the quantum geometric tensor and those introduced by Provost and Vallee \cite{Provost}, as well as Zanardi, Giorda, and Cozzini \cite{Zanardi}. As a matter of fact, the results obtained by analyzing every system under the path integral formalism  are the same as those coming from the other methods, which supports the theoretical validity of our proof.

Through examples we emphasize the geometric properties of the quantum metric tensor by computing the Ricci tensor and scalar curvature for each of them. This allows us to describe the parameter space of our systems accurately. For the general harmonic oscillator with a linear term, we find that even at low energy levels, the curvature of the parameter space is not flat. However, as the energy level increases, it becomes so. Furthermore, we consider a general Gaussian system with an arbitrary  dependency on two parameters and compute the associated quantum metric tensor, finding that its scalar curvature is $R=-4$, independently of the particular functional dependence of wave function on the parameters considered.

Regarding the symmetrically coupled harmonic oscillators, we notice a quantum phase transition when $\omega_1=0$ or $\omega_2=0$, which represents that any of our normal modes has become a free particle. However, we find that this system has a null Riemann tensor, and therefore its associated parameter space  is flat; by solving Beltrami's equation we found the coordinate system where this becomes explicit. Additionally, using the phase space contribution to the generalized quantum
geometric tensor we obtained the purity and von Neumann entropy $S$ for the ground state, finding that $S \to \infty$ as $k_0 \to 0$ or $k_1 \to \infty$. Furthermore, using the reduced quantum covariance matrix as a metric of parameter space we noticed that it's scalar curvature contains information related to the purity and the entropy. However, it has a different behavior from those two quantities in the limit when $k_1\to 0$. This characteristic must be studied in more detail for other entangled systems. Another feature of this curvature is that it explicitly shows the quantum phase transition observed in the determinant of the quantum metric tensor and, in general, keeps a nontrivial dependence on the quantum numbers in contrast with the quantum metric tensor's scalar curvature that in this case vanishes. 

The symmetrically coupled and linearly coupled harmonic oscillators provide examples where the scalar curvature associated with the phase-space contribution of the generalized quantum metric tensor, when regarded as a parameter space metric, contains valuable information about particle entanglement. We think that this quantity merits further investigation in this context.

In some cases, the path integral approach to the quantum geometric tensor could be more straightforward than other approaches, by incorporating in a natural way perturbation theory. This feature could be particularly useful in the study of the parameter space or quantum entanglement of non-exact systems \cite{AlvarezVergara-2019}, and paves the way for the calculation of the Green functions of our system via Feynman diagrams. Finally, it is worth pointing out that this formalism can be extended to quantum field theory by utilizing the LSZ relations~\cite{Srednicki1993}. 

\ack

This work was partially supported by DGAPA-PAPIIT Grant No. IN105422. D.G. acknowledges the financial support under Project SIP-20230323 of IPN. S.B.J. acknowledges the stipend Ayudante de Investigaci\'on of SNI.

\section*{References}
\bibliography{References}
\end{document}